\documentclass[structabstract]{aa}  
\usepackage{graphicx}
\usepackage{lscape}
\usepackage{amsmath}
\usepackage{txfonts}
\usepackage{url}
\usepackage{natbib}
\usepackage{xcolor}

\begin{document}
%
%   \title{Impact of enhanced mass loss during the red supergiant phase}
%   \title{Red supergiants as constraints and challenges for stellar evolution: impact of mass loss, rotation and metallicity}
 \title{The flux-weighted gravity-luminosity relationship of blue supergiant stars as a constraint for stellar evolution}
 
   \author{Georges Meynet\inst{1}, Rolf-Peter Kudritzki\inst{2, 3}, and Cyril Georgy\inst{4}
}

   \authorrunning{Meynet et al.}

 \institute{Geneva Observatory, University of Geneva, Maillettes 51, CH-1290 Sauverny, Switzerland
\and   Institute for Astronomy, University of Hawaii, 2680 Woodlawn Drive, Honolulu, HI 96822, USA  
 \and University Observatory Munich, Scheinerstr. 1, D-81679 Munich, Germany
\and Astrophysics, Lennard-Jones Laboratories, EPSAM, Keele University, Staffordshire ST5 5BG, UK
}

   \date{Received ; accepted }

% \abstract{}{}{}{}{} 
% 5 {} token are mandatory
 \abstract
  % context heading (optional) leave it empty if necessary  
   {
   The flux-weighted gravity-luminosity relationship (FGLR) of blue supergiant stars (BSG) links their absolute magnitude to the spectroscopically determined 
   flux-weighted gravity $\log\,g/T_\text{eff }^4$. BSG are the brightest stars in the universe at visual light and the application of the FGLR has become a powerful tool to determine extragalactic 
   distances.
   } 
   {
   Observationally, the FGLR is a tight relationship with only small scatter. It is, therefore, ideal to be used as a constraint for stellar evolution models. The goal of this work is
   to investigate whether stellar evolution can reproduce the observed FGLR and to develop an improved foundation of the FGLR as an extragalactic distance indicator.
   }
   {
   We use different grids of stellar models for initial masses between $9$ and $40\,M_\odot$, for metallicities between $Z=0.002$ and $0.014$, with and without rotation, computed with various mass loss 
   rates during the red supergiant phase. For each of these models we discuss the details of post-main sequence evolution and construct theoretical FGLRs by means of population synthesis
   models which we then compare with the observed FGLR.
   } 
   {
   In general, the stellar evolution model FGLRs agree reasonably well with the observed one. There are, however, differences between the models, in particular with regard to the shape and width 
   (scatter) in the flux-weighted gravity-luminosity plane.  The best agreement is obtained with models which include the effects of rotation and
   assume that the large majority, if not all the observed BSG evolve towards the red supergiant phase and only a few are evolving back from this stage. The effects of metallicity on the shape and 
   scatter of the FGLR are small.
   }
 {
  The shape, scatter and metallicity dependence of the observed FGLR are well explained by stellar evolution models. This provides a solid theoretical foundation for the use of this 
   relationship as a robust extragalactic distance indicator.  
  }
%   {Changing the mass loss rates during the red supergiant phase has many consequences for the evolution of massive stars and will have a consequence also on the
%   expected luminosity function of red suoergiant stars. Higher the mass loss, steeper will be the decline of the luminosity function of the red supergiants in a constant star formation %system.}
 

 \maketitle
%==================================================================
%__________________________________________________________________

\section{Introduction}

\begin{figure*}
\centering
\includegraphics[width=.5\textwidth, angle=0]{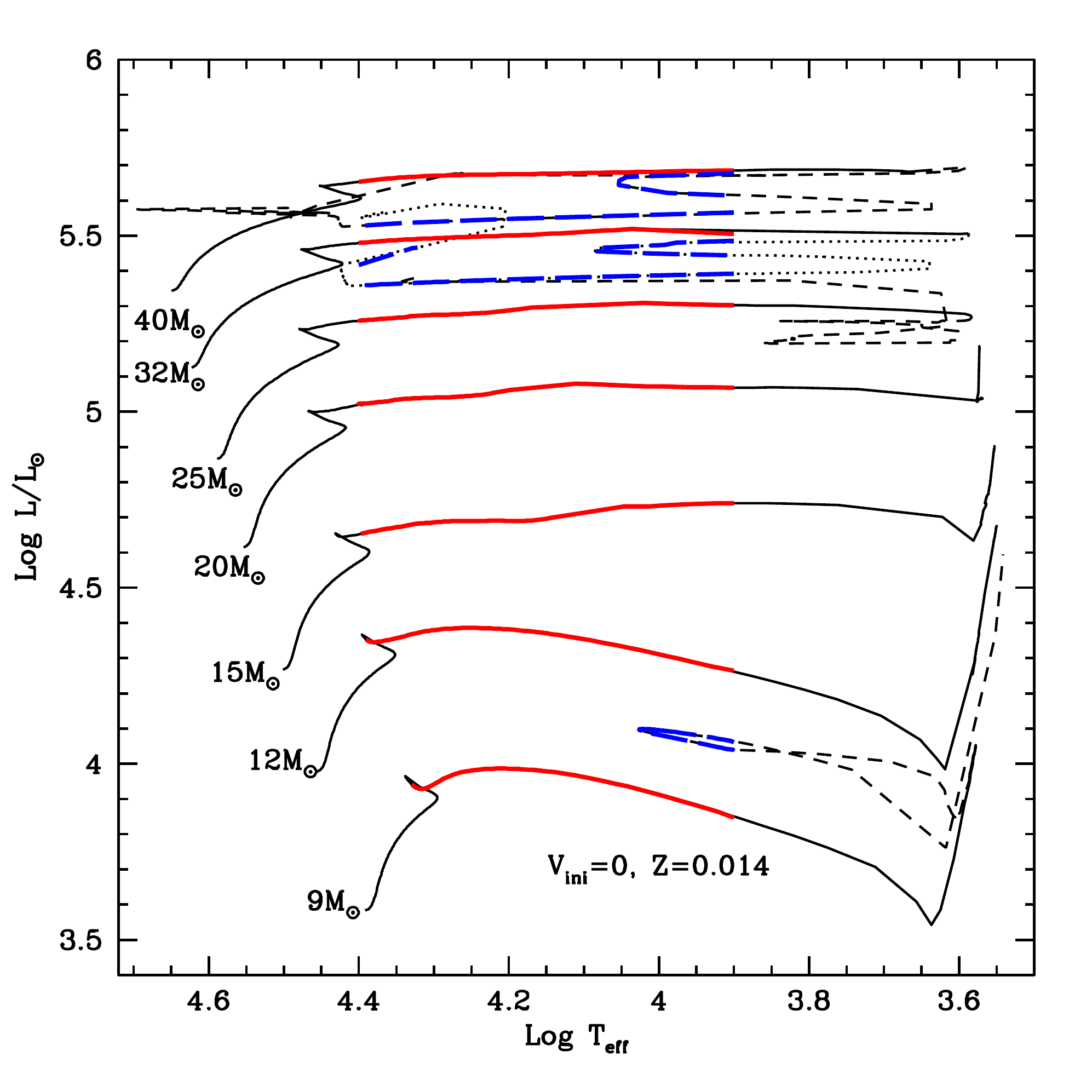}\includegraphics[width=.5\textwidth, angle=0]{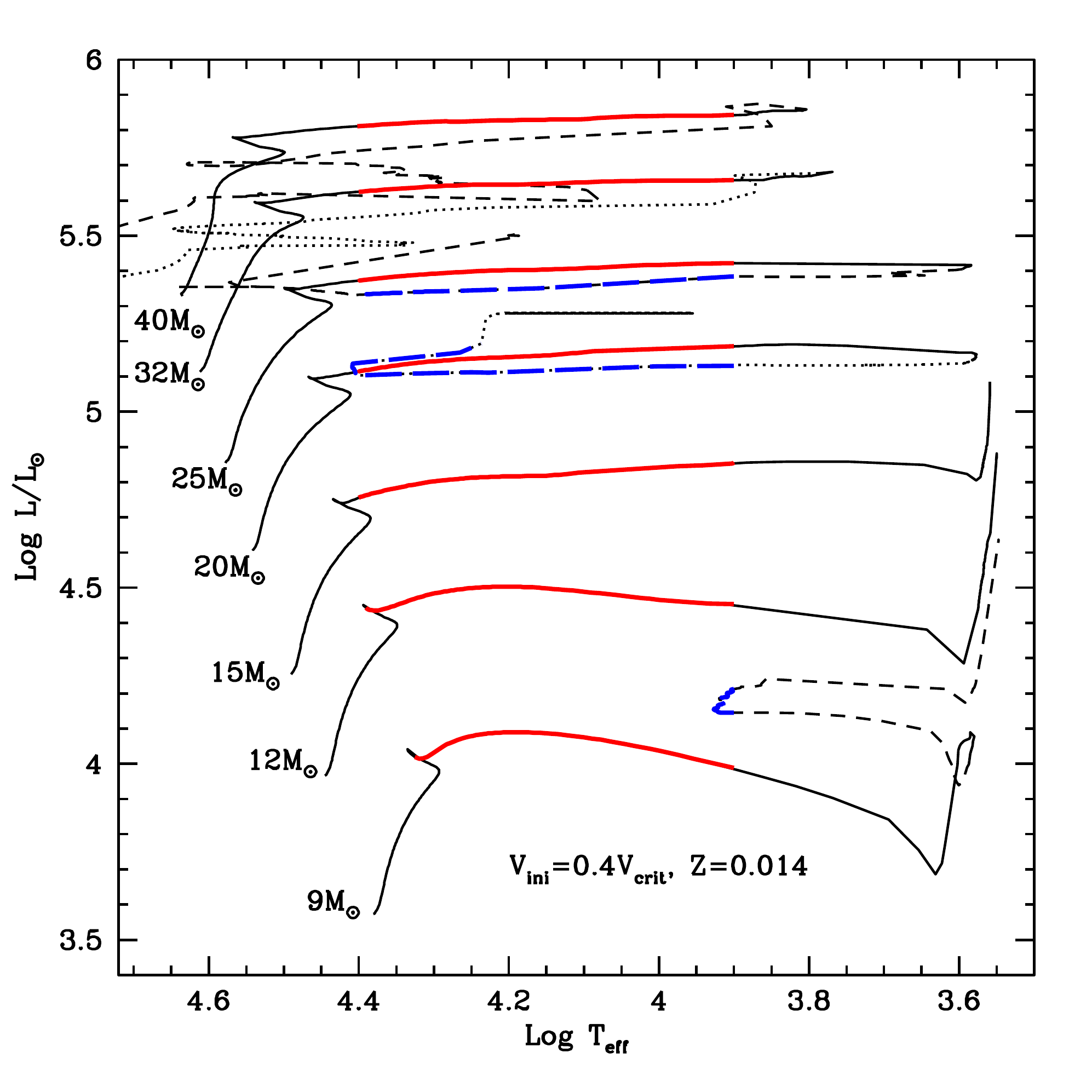}
   \caption{Evolutionary tracks in the Herstzsprung-Russel diagram calculated for solar metallicity $Z=0.014$ for initial masses $9$, $12$, $15$, $20$, $25$, $32$ and $40\,M_\odot$. 
   The continuous lines show the track until the red supergiant phase. After the red supergiant stage, either dashed or dotted lines are used.
   The BSG phase
   is highlighted in red for the evolution towards the RSG phase (group 1, see text) and in blue (dashed) for the evolution back after the RSG stage (group 2). 
   The models are from \citet{Ekstrom2012}.
   {\it Left panel: } Models without the effects of stellar rotation. {\it Right panel: }  Models including rotation (see text).}
      \label{hrd}
\end{figure*}

Blue supergiants (BSG) are massive stars in the mass range between $10$ to $40\,M_{\odot}$ in the short-lived 
evolutionary phase when they cross the Hertzsprung-Russel diagram (HRD) between the hydrogen main sequence 
(MS) and the red supergiant (RSG) stage (see Fig.~\ref{hrd}).  BSGs may also evolve from a RSG stage as a result of very strong mass-loss.
With effective temperatures in a range between $20000\,\text{K}$ to $8000\,\text{K}$ the maximum of the energy distribution of these extremely luminous 
objects shifts towards 
visible wavelengths, which makes them the brightest stars in the universe at visual light. With absolute visual magnitudes $\text{M}_\text{V} \cong -9.5\,\text{mag}$, a single BSG can be as bright as a globular cluster or 
a dwarf galaxy. Because of their brightness they are ideal objects to study the chemical composition of the young stellar population in galaxies far beyond the local group and to determine 
extragalactic distances using the tools of quantitative stellar spectral analysis (see, for instance, \citealt{Kud2014} and references therein). At the same time, BSGs are also extremely useful to constrain models
of stellar evolution either through the investigation of their chemical surface composition revealing the presence of rotationally or convectively induced mixing \citep[see e.g.][]{GiLa92, 
Hunteretal08a, Hunter08,  Przybilla10, Urbaneja11, FiPr12, McEvoy14, Maeder14}, or through the number ratios of BSG, RSG and MS stars \citep{Dohm02, Eggenberger2002} or the study of their pulsational
properties \citep{Saio2013}.

Guided by the theory of stellar evolution, which predicts that BSG cross the HRD at approximately constant luminosity and constant mass, \citet{Kudritzki2003}
\citep[see also][]{Kud2008} 
have detected a tight relationship between absolute bolometric magnitude $\text{M}_\text{bol}$  and the
spectroscopically determined flux-weighted gravity $\log{g/T_\text{eff,4}^4}$, the flux-weighted gravity - luminosity relationship (FGLR) ($g$ is the gravity and $T_\text{eff,4}$, the effective temperature in units of $10^4\,\text{K}$). 
The application of the FGLR using low resolution optical spectra of individual BSGs in galaxies has a great potential as a precision distance indicator for the extragalactic 
distance scale. The spectroscopic determination of stellar temperature, gravity and metallicity combined with multi-color photometry allows for an accurate determination of reddening, reddening law 
and extinction of each individual BSG, thus, avoiding the extinction and metallicity induced uncertainties encountered with Cepheid stars. The de-reddened apparent magnitude in conjunction with
the absolute magnitude obtained from the FGLR and with the spectroscopically determined flux-weighted gravity $\log{g/T_\text{eff,4}^4}$ can then be used to determine accurate distances. Over the last 
years, this new method has been successfully applied to galaxies out to a distance of $7\,\text{Mpc}$ (\citealt{Urba2008}, \citealt{U2009}, \citealt{Kud2012}, \citealt{Kud2013}, \citealt{Hosek2014}, 
\citealt{Kud2014}).

Besides the potential as distance indicator the observed FGLR of BSG can also be used to constrain the stellar evolution models. This is purpose of the work presented in this paper. 
Using stellar evolutionary tracks we shall address mainly three questions:

1) The population of blue supergiants is made from stars starting their evolution with different initial rotational velocities. 
   This leads to the question, how important are the effects of rotation for the post-main sequence evolution?

2) How does stellar metallicity affect the post-main sequence evolution and the FGLR?

3) As indicated in Fig.~\ref{hrd}, BSG can be in two different evolutionary phases, evolving towards the red supergiant stage (RSG) or back to hotter temperatures from the RSG stage. 
   The fact that some core collapse supernovae have a yellow or even a blue progenitor (e.g. see table 6 in \citealt{Meynet14}) indicates that the latter phase does indeed occur in nature and is not just a theoretical artefact. Moreover, some pulsation properties of blue supergiants are much better accounted for if the objects are in a post RSG stage 
   \citep{Saio2013}. This leads to the immediate question about the fraction of post-RSG objects among the BSG stars and the reliability of post-RSG evolutionary tracks.

The stellar models used to investigate these questions are briefly presented in Section 2. 
Section 3 describes the stellar evolution background for the existence of the FGLR. 
Section 4 compares theoretical FGLRs with the empirical one. We will demonstrate
that indeed some stellar evolution models can be ruled out as being representative of the bulk population.
Section 5 shows results of population synthesis models providing an alternative way to compare predictions of the theory with the observed FGLR. 
Section 6 contains the main conclusions and discusses some aspects of future works.

\section{The stellar models}

The stellar evolution models used for this investigation are from the solar metallicity grid of \citet{Ekstrom2012}, the grid of \citet{Z002G2013} for the metallicity $Z=0.002$, and from an unpublished grid
of stellar models for the metallicity $Z=0.006$ (Eggenberger et al., in preparation). Moreover, we use the models with an enhanced mass loss rate model
during the RSG phase recently published by \citet{Meynet14}.

The models are computed using the Schwarzschild criteria for convection with core overshooting. The core extension due to overshooting is taken equal to $10\%$ the pressure scale height estimated at 
the Schwarzschild core boundary. Non adiabatic convection is considered in the outer convective zone with a mixing length scale equals to 1.6 times the local pressure scale height. 
In rotating models, the shear turbulence coefficient is taken from \citet{Maeder1997}, while the horizontal turbulence and the effective diffusion coefficients are those from \citet{Zahn1992}.
Initial velocities on the ZAMS have been taken equal to 40\% the critical velocity. Typically for the models at $Z$=0.014, this corresponds to initial equatorial velocities between 248 and 314 km s$^{-1}$ for initial masses between 9 and 40 M$_\odot$.

The mass-loss prescription for the hot part of the evolutionary tracks is that of  \citet{dejager1988} for the initial masses $9$ and $15\,M_\odot$ and for $\log(T_\text{eff}/\text{K}) > 3.7$. 
For $\log(T_\text{eff}/\text{K}) < 3.7$, we use a fit to the data by \citet{Syl1998}
and \citet{vanLoon1999} as suggested by \citet{Crowther2001}. Above $15\,M_\odot$, the prescription given by \citet{Vink2001} is used on the MS phase as long as $\log(T_\text{eff}/\text{K}) > 3.9$, 
the recipe from \citet{dejager1988} is used
for the non red supergiant phase. For $\log(T_\text{eff}/\text{K}) < 3.7$, the prescription is the same as for lower initial mass stars.
The effects of rotation on the mass-loss rates are accounted for as in \citet{mm6}. 
Note that these effects are quite negligible for the rotation rates considered in this work.

As explained in \citet{Ekstrom2012}, for massive stars ($>15\,M_\odot$) in the red supergiant phase, some parts of the most external layers of the stellar envelope might exceed the Eddington luminosity of the star: $L_\text{Edd} = 4\pi cGM/\kappa$ (with $\kappa$ the opacity). This is due to the opacity peak produced by the variation of the ionization level of hydrogen beneath the surface of the star. 
We account for this phenomenon by increasing the mass-loss rate of the star (computed as described above) by a factor of $3$. Once the supra-Eddington layers disappear, later during the evolution, we come back to the usual mass-loss rate. 

For the enhanced mass-loss rate models, we just multiply our standard mass-loss rates by a factor 10 or 25 during the whole period when the stars are in the RSG-phase.
(We consider a star as a RSG when its effective temperature ($T_\text{eff}$), as estimated by the Geneva code, is $\log T_\text{eff}(/\text{K}) < 3.7$.) 
Note that when the enhanced mass loss rates are used, we do not account for the effect of the supra-Eddington layers as described in  \citet{Ekstrom2012}. This means that the enhancement of the mass-loss rates with respect to those used in \citet{Ekstrom2012} are actually a little less than the factor 10 and 25 \citep[see][for a more detailed discussion]{Meynet14}.

\section{Some theoretical considerations about the flux-weighted gravity-luminosity}

As shown in \citet{Kud2008}, assuming a mass-luminosity relation of the type
$\log L/L_\odot =\alpha \log M/M_\odot + b,$
and the definition of $\text{M}_\text{bol}$, $\text{M}_\text{bol}=-2.5 \log L/L_\odot +4.75,$
one obtains the FGLR as a linear relationship between $\text{M}_\text{bol}$ and $\log g/T_\text{eff,4}$, 

\begin{align}
\text{M}_\text{bol}=&2.5 \frac{\alpha}{\alpha-1}\left[\log\frac{g}{T_\text{eff,4}^4}-16\right]\notag\\
&-2.5\frac{\alpha}{\alpha-1}\left[\log(4\pi \sigma G)+\log\frac{M_\odot}{L_\odot}\right]\notag\\
&+2.5\frac{b}{\alpha-1}+4.75.\label{eqone}
\end{align}

The symbols have their usual meaning. We see that the slope of the mass-luminosity relation $\alpha$ affects the slope of the FGLR. When $\alpha$ decreases, the FGLR slope increases.
At the same time $\text{M}_\text{bol}$ decreases (i.e. it becomes more negative). We also see that $\text{M}_\text{bol}$ increases with $b$.

\begin{figure*}
\centering
\includegraphics[width=.5\textwidth, angle=0]{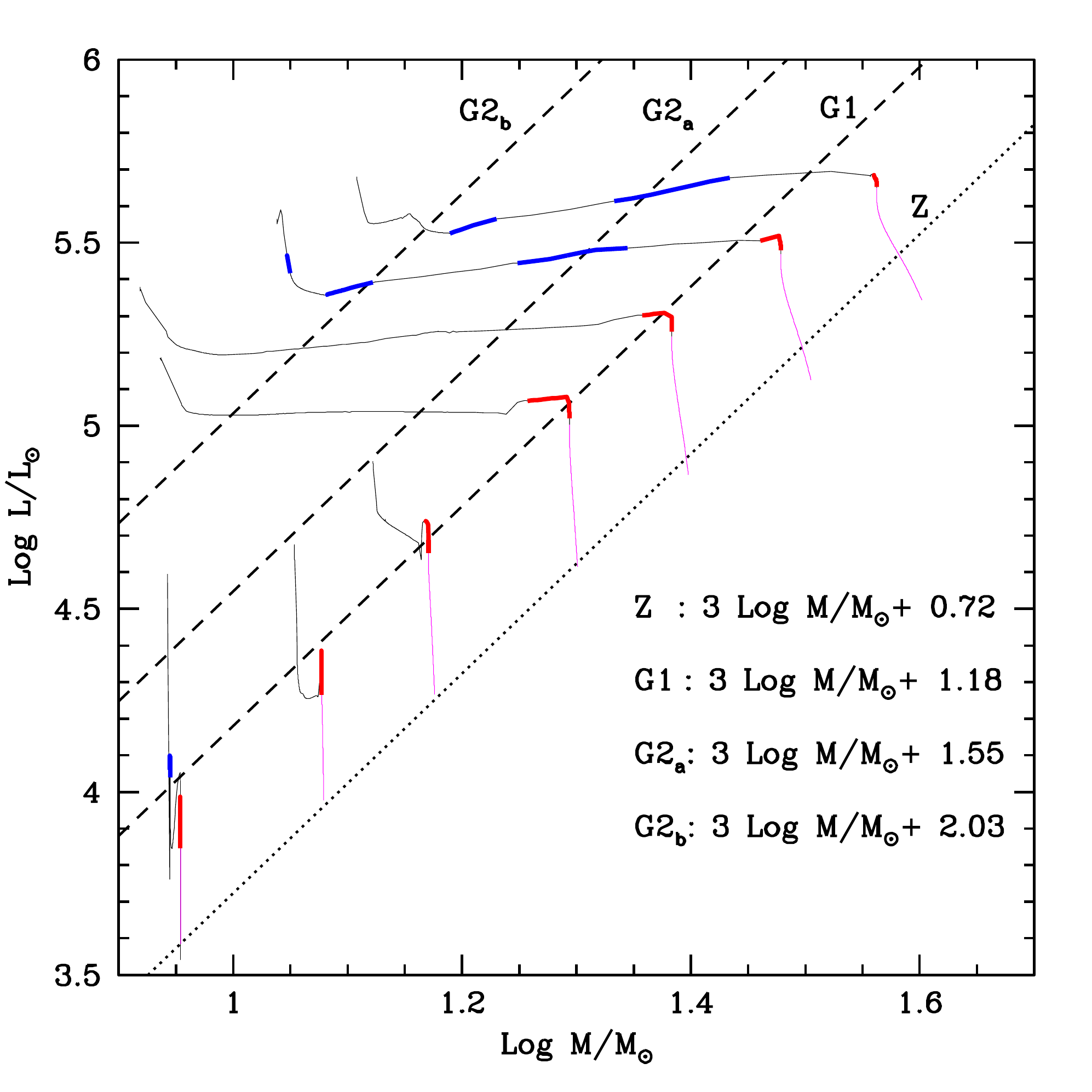}\includegraphics[width=.5\textwidth, angle=0]{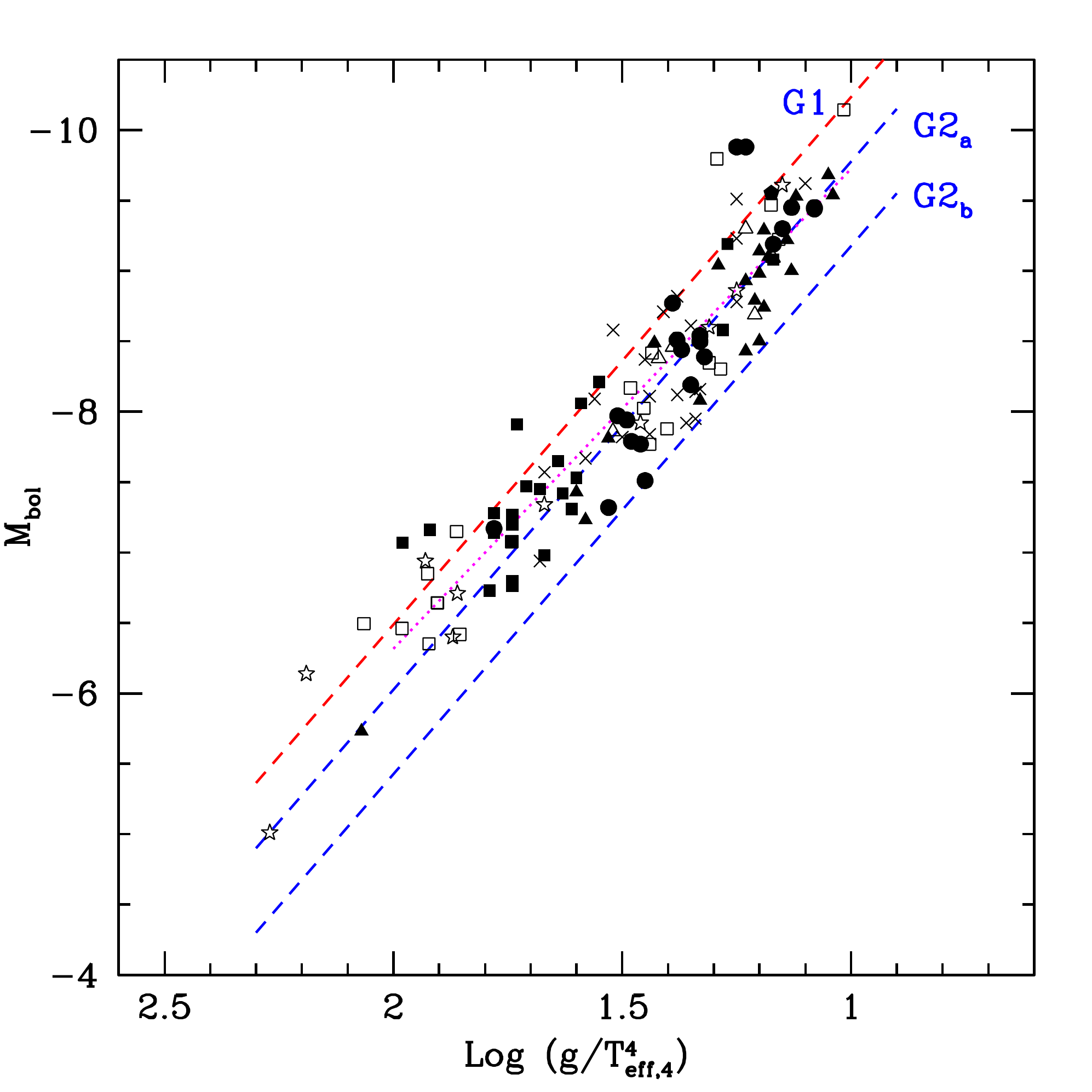}
   \caption{{\it Left panel: } Evolutionary tracks at solar metallicity without rotation in the ($\log M/M_\odot$, $\log L/L_\odot$)-plane. $M$ is the actual mass of the star during the evolution.
   The dashed lines are mass-luminosity relations with $\alpha=3$ and $b=0.72316$, $1.17955$, $1.54727$ and $2.03346$, respectively (bottom-up).
   The lower vertical or nearly vertical (magenta) part of the track corresponds to the Main-Sequence phase, while the red and blue parts show the group 1 and group 2 phases before and after the red supergiant stage.
   {\it Right panel: }  Flux-weighted gravity-luminosity relations obtained from Eq.~\ref{eqone} and the mass-luminosity relations label G1, G2$_\text{a}$ and G2$_\text{b}$ on the left panel.
    The points represent observations of individual BSG. The black squares are the values for NGC 300 given by
  \citet{Kud2008}. The full triangles are for supergiants in other galaxies studied by \citet{Kud2008b}. Full circles and crosses are blue supergiants in M33 \citep{U2009} and M81 
\citep{Kud2012}, respectively. The empty stars and empty squares show the data for the metal poor Local Group galaxies WLM \citep{Urba2008} and NGC~3109 by \citep{Hosek2014}. The empty triangles and 
the full pentagons correspond to BSG in NGC 3621 \citep{Kud2014} and in NGC 4258 \citep{Kud2013}, respectively. The dotted magenta line represents the FGLR calibration obtained by \citet{Kud2008}.}
      \label{lm}
\end{figure*}

Are the blue supergiants following a well defined mass-luminosity relation?
To investigate this point we have constructed Fig.~\ref{lm}. In the left panel, the non-rotating  solar metallicity evolutionary tracks are shown in a luminosity-mass diagram. 
The evolutionary tracks start vertically from the bottom but then evolve towards the left with decreasing mass due to mass loss. The vertical phase at the very beginning coincides with 
the main-sequence phase, where the mass loss rates are very modest. 
After the Main-Sequence phase, the extension to the left is a consequence of mass loss, while the luminosity keeps a nearly constant value. 
The evolutionary stage of blue supergiants defined here as 
post MS stars with effective temperatures in the range between $\log T_\text{eff} = 3.9$ and $4.4$ is indicated as heavy sections in red, when the blue supergiants are on their
first crossing towards the RSG phase (group 1 BSG),  and in blue, when they come back after the red supergiant phase (group 2 BSG). Note that we exclude late phase stellar evolution models
with extremely high helium surface abundance (corresponding to a hydrogen surface mass fraction $X_\text{surf}$ smaller than 0.3) from the BSG group 2. Such high helium abundances are not found
by the quantitative spectral analyses of normal BSG \citep{FiPr12} and may belong to objects with Wolf-Rayet-star or Luminous Blue Variable spectral characteristics.

We also overplot some mass-luminosity relations with constant and parallel slopes.
This is an approximation made for simplicity. In reality the mass luminosity relation flattens at the high mass end, an effect which is caused
by the increasing importance of radiation pressure in the stellar interior \citep[see e.g.][]{Maeder09,Yusof2013}. However, in order to demonstrate the most important effects the simple approximation of 
constant slope is sufficient.

Relation Z represents the ZAMS evolutionary stages. Relation G1 connects the group 1 BSG. We note, however, that the blue-loop core He-burning phase of the $9\,M_\odot$ track produces group 2 BSG 
with very similar mass and luminosity. This comes from the fact that mass-loss is not important for this relatively low mass. The relations G2$_\text{a}$ and G2$_\text{b}$ represent the two phases 
at higher mass of the group 2 BSG (see Fig.~\ref{hrd}). 

In the next step, we use the different  mass-luminosity relations connecting the various groups of blue supergiants
to construct  flux-weighted gravity-luminosity relations. We obtain the three lines plotted in the right panel of Fig.~\ref{lm}.
The sample of observed blue supergiants used to obtain the empirical flux-weighted gravity-luminosity relation is also shown (see the caption for the references of these observations).

We see that assuming the luminosity scaling with the third power of the mass provides a good fit of the average slope of the empirical relation.
We also see that the empirical relation deduced by \citet{Kud2008} (the magenta dotted line) has a slope very close to the simple theoretical
relations. In addition, we note that the flux-weighted gravity-luminosity relations of group 2 blue supergiants are shifted downwards. The shift of G2$_\text{b}$ amounts to almost one bolometric magnitude. This shift is not so much due to
a decrease in luminosity (see the left panel of Fig.~\ref{lm}), but to a decrease in  $g/T_\text{eff}^4$ (rightward shift in the right panel of Fig.~\ref{lm}).
This can be understood in the following way. $g/T_\text{eff}^4$ is proportional to $M/L$ and, while $L$ remains more or less constant between the evolutionary phases of group 1 and 2,
the mass is decreased due to strong mass-loss
during the RSG stage, where stars may lose between 25-40\% of their initial mass \citep[see e.g. Fig. 3 in][]{Meynet14}.

From this brief discussion, we can summarize the following main points: there is a one-to-one connection between mass-luminosity relations and flux-weighted gravity-luminosity relations.
The group 1 and group 2 blue supergiants follow different mass-luminosity relations and, thus, different flux-weighted gravity-luminosity relations.
Group 1 appears to form the upper envelope of the observed points in the $\text{M}_\text{bol}$ versus $\log (g/T_\text{eff,4}^4)$ plane, while the group 2 cover
the averaged positions of the observed points, as well as their lower envelope. 

At first sight, one could conclude from these first comparisons that the observed scatter is compatible with the existence of group 1 and group 2 blue supergiants as predicted by the
models. However, as we shall see below, this is not correct. Actually, this simple comparison suffers from many weaknesses:
First, while we have plotted the relations G2$_\text{a}$ and G2$_\text{b}$ over the whole flux-weighted gravity range, the models predict the corresponding group 2 stars only for the highest 
masses and luminosities. Second, we need to consider the time scales during the evolutionary phases corresponding to group 1 and the various group 2 phases in order to assess how, according
to the models, the blue supergiants should be distributed along the different relations. 
Third, the scatter of the observed FGLR is, of course, not only affected by the presence of various types of blue supergiants but also by the observational uncertainties in the
determination of bolometric magnitude and flux-weighted gravity. In order to disentangle these effects for a more precise comparison
we need to construct the flux-weighted gravity-luminosity relations as they result from the tracks corresponding exactly during the group 1 and 2 BSG phases including the information of the 
relative time scales to blue supergiants. This is the topic of the next sections. 

\section{Comparisons with observations}

In the following we discuss a sequence of evolutionary models in the ($\log g/T_\text{eff,4}^4$, $\text{M}_\text{bol}$)-plane of the FGLR and compare with the observed distribution of BSG. The plots show only
post-main sequence evolution and again highlight the group 1 and 2 phases in red and blue as before. We will start with models at solar metallicity assuming standard rates of mass-loss. We will then 
consider the effects of enhancing mass-loss in the RSG phase and will finish this section investigating the effects of reduced metallicity.

\subsection{Models for solar metallicity with standard mass loss rates}

\begin{figure*}
\centering
\includegraphics[width=.37\textwidth, angle=-90]{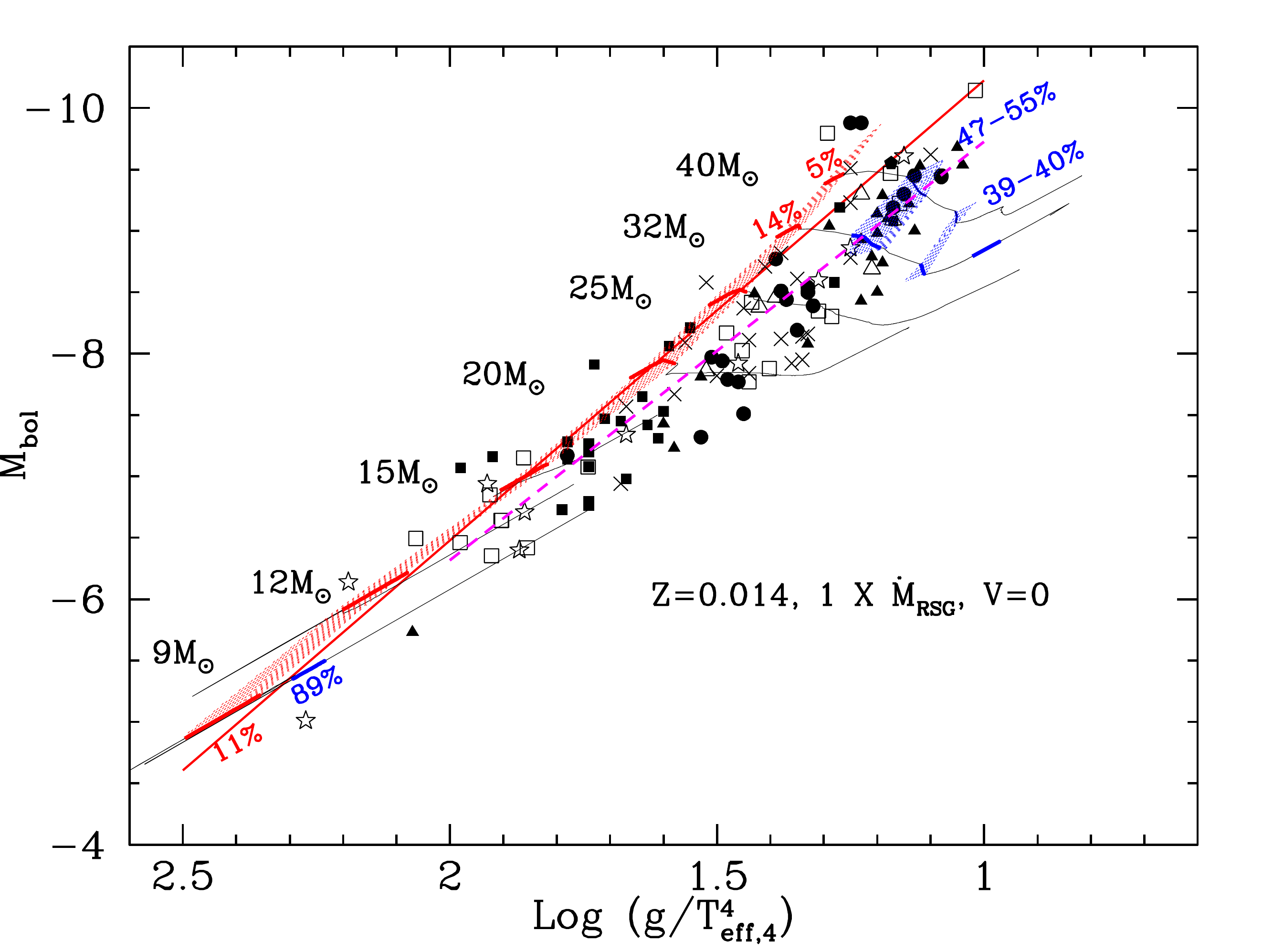} \includegraphics[width=.37\textwidth, angle=-90]{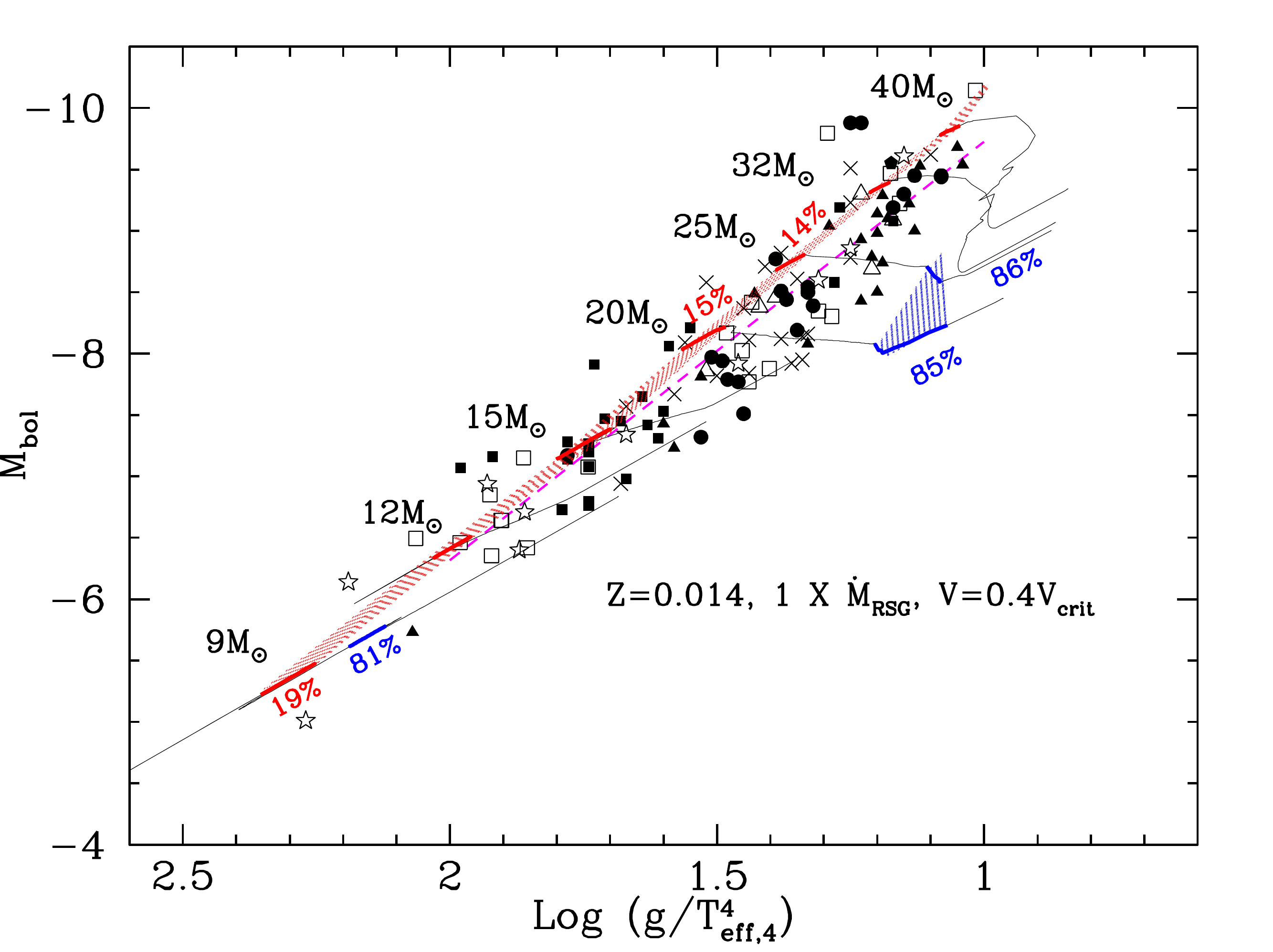}
   \caption{Post-main sequence evolutionary tracks labelled by the original mass on the ZAMS in the ($\log g/T_\text{eff,4}^4$, $\text{M}_\text{bol}$)-plane for $Z=0.014$ and standard mass loss rates in the 
   RSG stage. The tracks are shown as continuous lines, which are then highlighted in heavy red or blue for the BSG group 1 and group 2 phases, respectively. The red and blue hatched areas connect
   the group 1 and group 2 phases of tracks with different mass. Relative fractions of the evolutionary times spent in the group 1 and 2 phases are also given in red and blue.
   {\it Left panel:} Evolutionary tracks without the effects of rotation. The straight red line corresponds to relation G1 in Fig.~\ref{lm}. The magenta dashed-line corresponds to the empirical relation obtained by \citet{Kud2008}. {\it Right panel:} Evolutionary tracks including rotation.}
      \label{fig14m1}
\end{figure*}

Fig.~\ref{fig14m1} shows $Z=0.014$ models calculated with standard mass-loss rates (see section 2). We begin with a discussion of non-rotating models.
The first important fact we note is that the heavy red group 1 phases of each track are not reduced to a single point as it would be if mass and luminosity were strictly constant. 
However, the group 1 segments corresponding to each track are very short and more or less parallel to each other. In consequence, the hatched area which connects these segments,
forms a narrow sequence, which represents the BSG group 1 FGLR for these models. At higher mass or luminosity the slope of this FGLR sequence becomes steeper. 
This is due to the fact, already noted in the previous section,  that the slope of the mass-luminosity flattens when the initial mass increases. We also see that
the sequence constitutes an upper envelope of the observed points in the upper luminosity range. This was already mentioned in \citet{Kud2008}. 
Taken at face value, this could indicate that group 1 does not constitute the bulk of the blue supergiant population at high luminosity.
This would be in line with the evolutionary time scales in the high mass range indicating, for instance, for the $32\,M_\odot$ model that
only about $14\%$ of the blue supergiants are group 1 BSG, while $47\%$ and $39\%$ would be in the two groups 2 regions.
Thus, in this luminosity range, group 2 blue supergiants could indeed reproduce the shape of the observed FGLR including its width at the high luminosity end.
For lower luminosities, however, the models appear to be in slight disagreement with the observations, because of the absence of group 2 BSG and the fact the group 1 sequence is at too high
bolometric magnitudes.

The situation changes with the models including the effects of rotation. First, the group 1 sequence shifts to the right towards slightly lower flux-weighted gravities. This is the effect of
the rotational mixing of helium and heavier elements, which decreases the stellar mass to luminosity ratio \citep[see e.g. the review by][and references therein]{2012RvMP}. The effect becomes more
efficient at higher stellar mass which leads to a significantly shallower slope of the group 1 FGLR sequence in better agreement with the observations. Second, the group 2 BSG extends towards lower
stellar masses. This results from stronger mass-losses during the RSG phase which comes from the fact that rotating models have higher luminosities. The higher luminosity and the reduced mass due 
to stronger mass-loss also causes a shift to the right for the group 2 objects.

As in the case of the non-rotating models, the evolutionary time scales indicate that most of the blue supergiants are expected to be group 2 BSG. This does not seem to agree with the observations. 
However, we have to add a few cautious remarks at that point:
The observed sample should results from stars with different initial velocities and, therefore, it may be that
most observed stars begin their evolution with rotation less than $40\%$ the critical velocity on the ZAMS.
Second, the observed blue supergiants displayed in Fig.~\ref{fig14m1} have different initial metallicities. In fact, slightly more than half of the observed stars (about $53\%$) have actually a metallicity lower than half
solar. As we shall see later, while the influence of metallicity on the group 1 blue supergiants in the
$\text{M}_\text{bol}$ versus $\log g/T_\text{eff,4}^4$ plane is very weak, the impact on the group 2 is in contrast quite important.
Third, the mass loss rates during the RSG stages are still very poorly known. Lower mass loss rates during the RSG
stage would shift the blue regions in Fig.~\ref{fig14m1} to the left making the positions of these stars more compatible with the observed scatter. At the same time, this also reduces the 
evolutionary time in the group 2 phase, since with lower mass-loss rates the stars spend more time in the RSG stage and after that evolve faster back to hotter temperatures.

To summarize the discussion of solar metallicity evolution models with standard mass-loss rates we note: 
\begin{itemize}
\item group 1 BSG form a well defined FGLR sequence and with a width much smaller than the one observed. 
Models with rotation agree better with the observations. 
\item The predicted location of group 2 BSG is in marginal agreement with the observations only for non-rotating tracks.
For tracks with rotation mass-loss in the RSG phase plays an important role and high mass-loss rates seem to lead to a disagreement with the observations.
\end{itemize}

The second point will be further investigated in the next subsection for the case of higher mass-loss rates.

\subsection{Models at solar metallicity with RSG enhanced mass loss rates}

\begin{figure*}
%: fig TcrcZ014_Mcc_rot.pdf
\centering
\includegraphics[width=.38\textwidth, angle=-90]{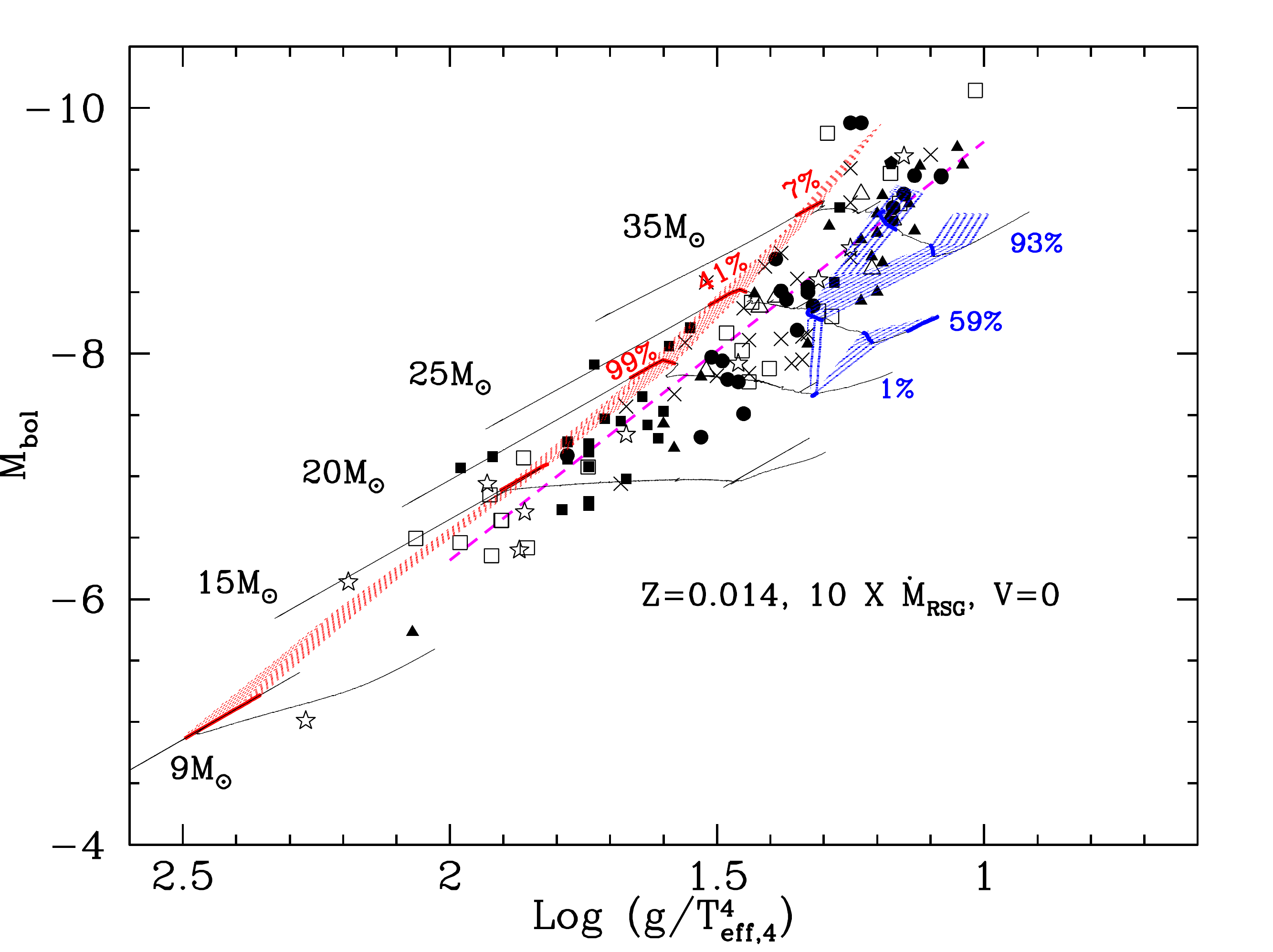}\includegraphics[width=.38\textwidth, angle=-90]{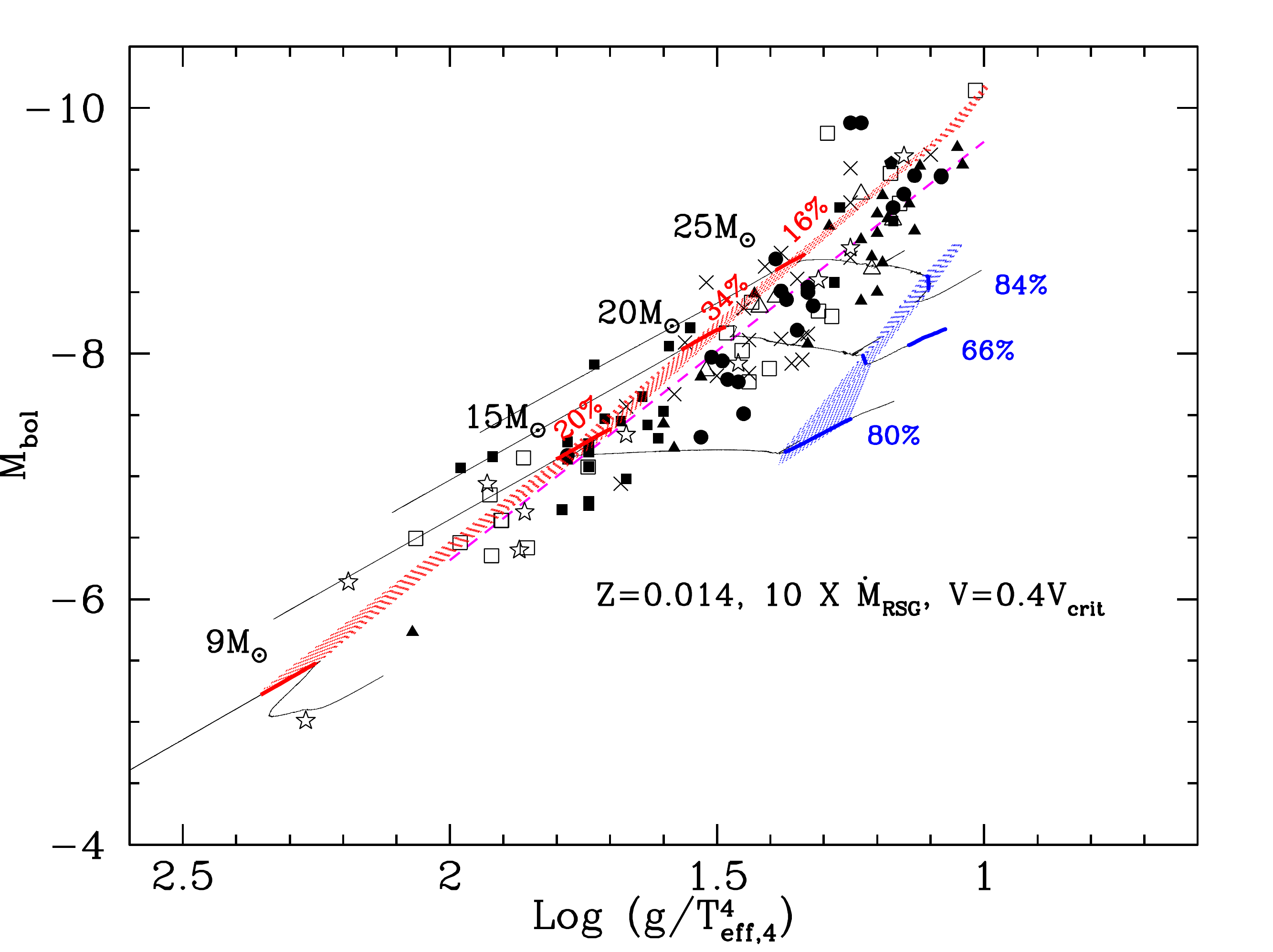}
\includegraphics[width=.38\textwidth, angle=-90]{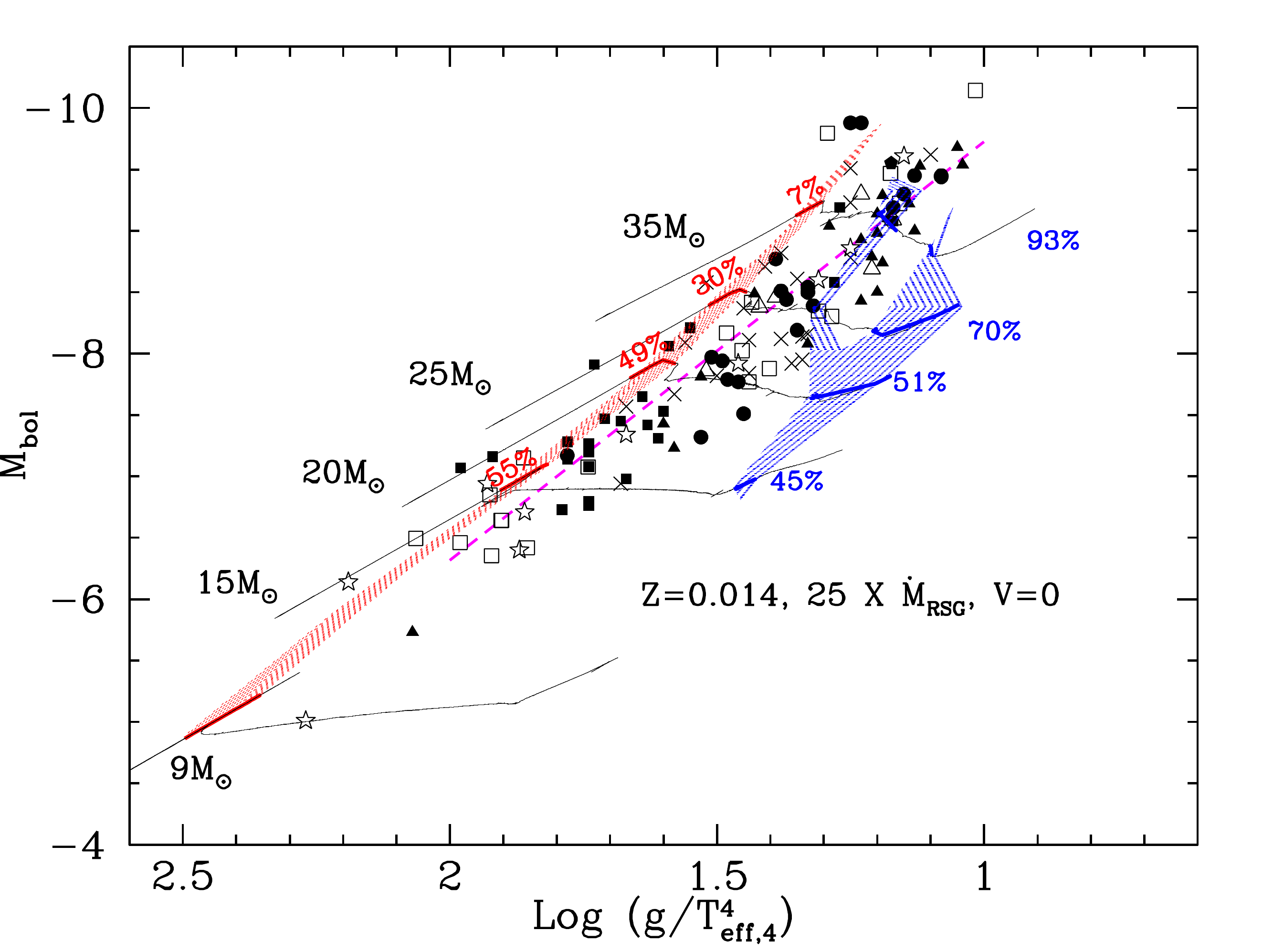}\includegraphics[width=.38\textwidth, angle=-90]{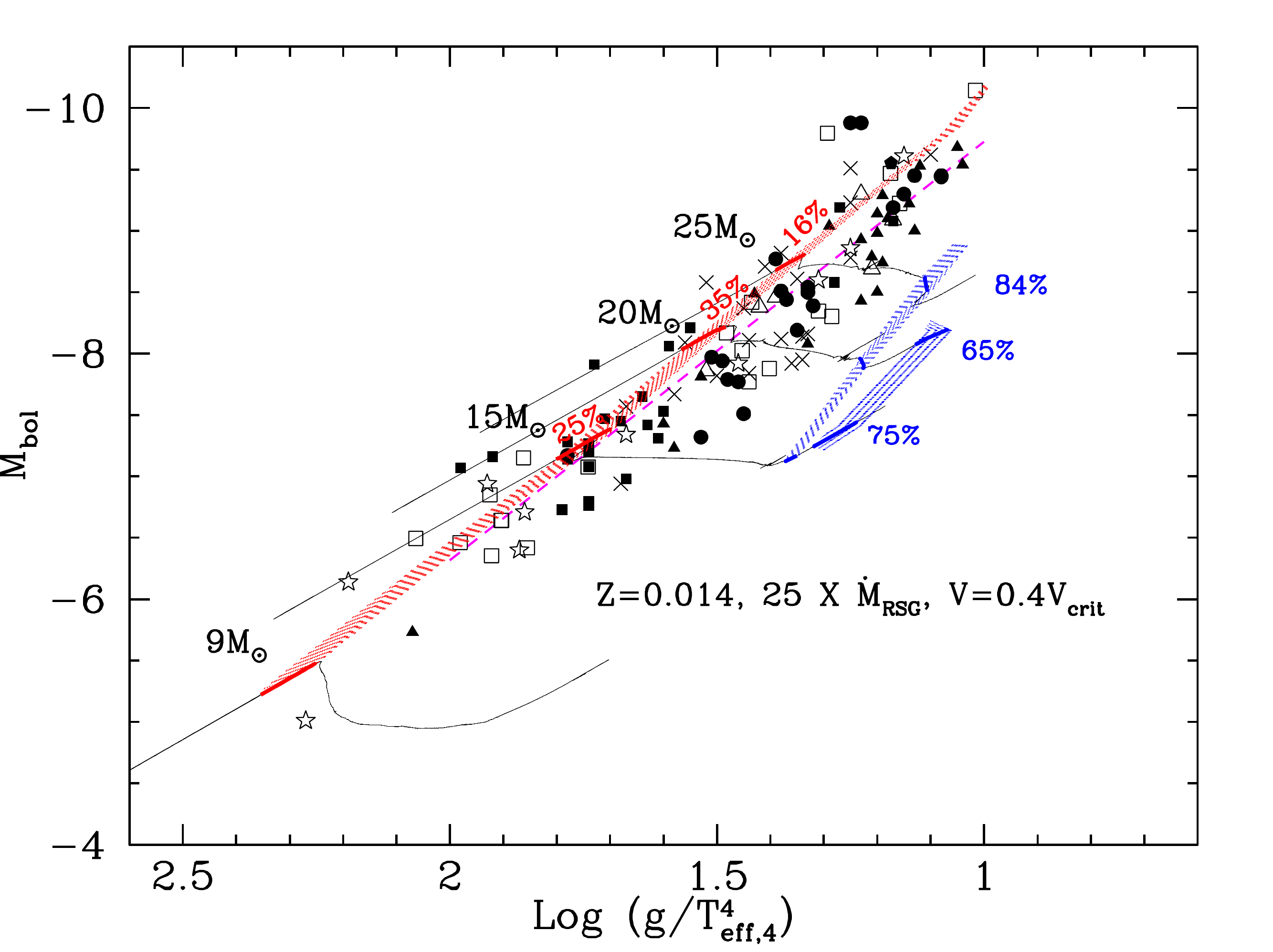}
   \caption{{\it Upper left panel:} Same as Fig.~\ref{fig14m1} for models without rotation but with RSG mass-loss rate enhanced by a factor 10 and for $Z=0.014$. 
    {\it Upper right panel: } Same as the upper left panel but for models with rotation.
   {\it Lower left panel: } Models without rotation but with RSG mass-loss rate enhanced by a factor 25. {\it Lower right panel: } Same as the lower left panel but for models with rotation}
      \label{figmdot}
\end{figure*}

\begin{figure*}
%: fig TcrcZ014_Mcc_rot.pdf
\centering
\includegraphics[width=.38\textwidth, angle=-90]{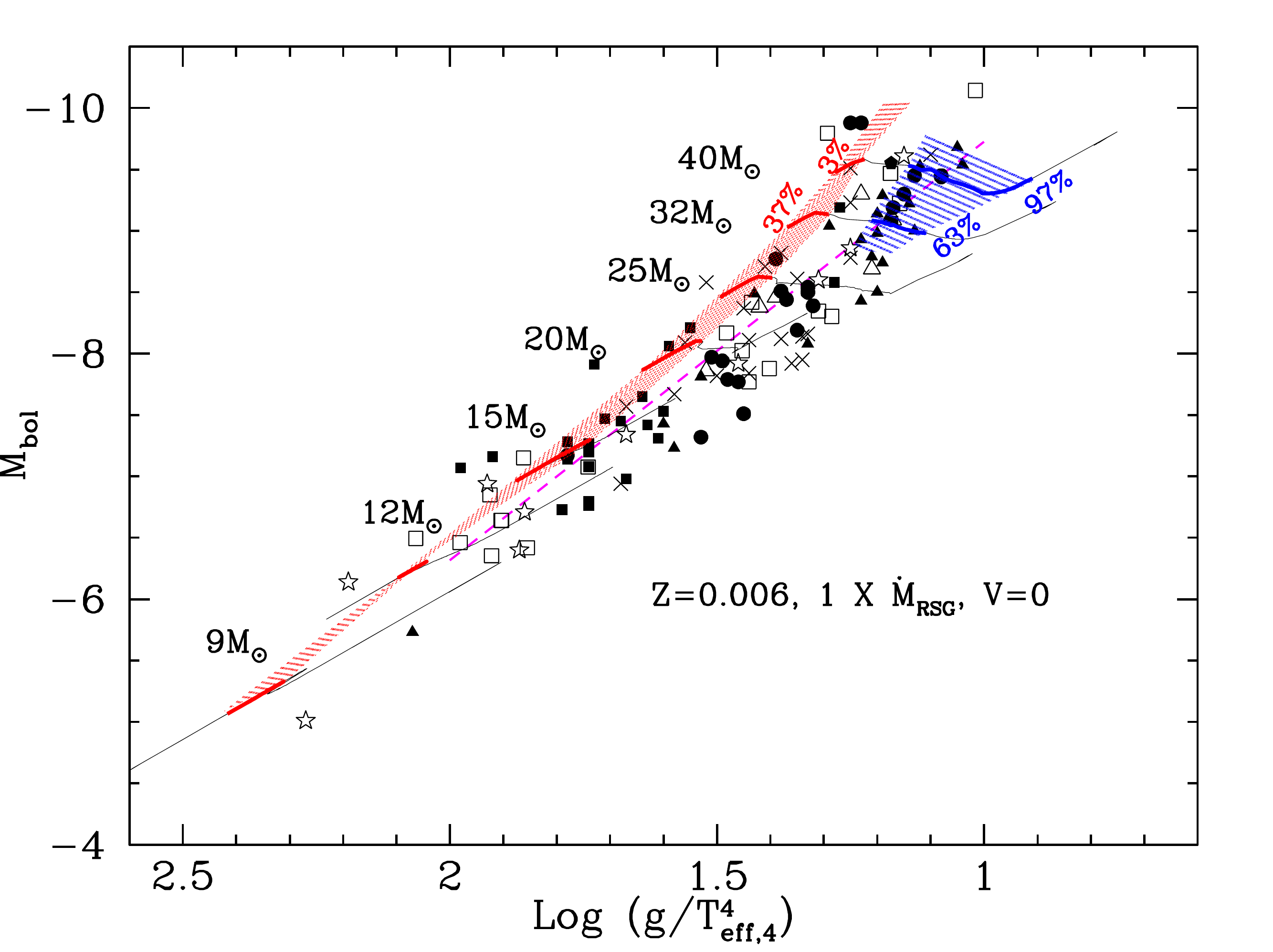}\includegraphics[width=.38\textwidth, angle=-90]{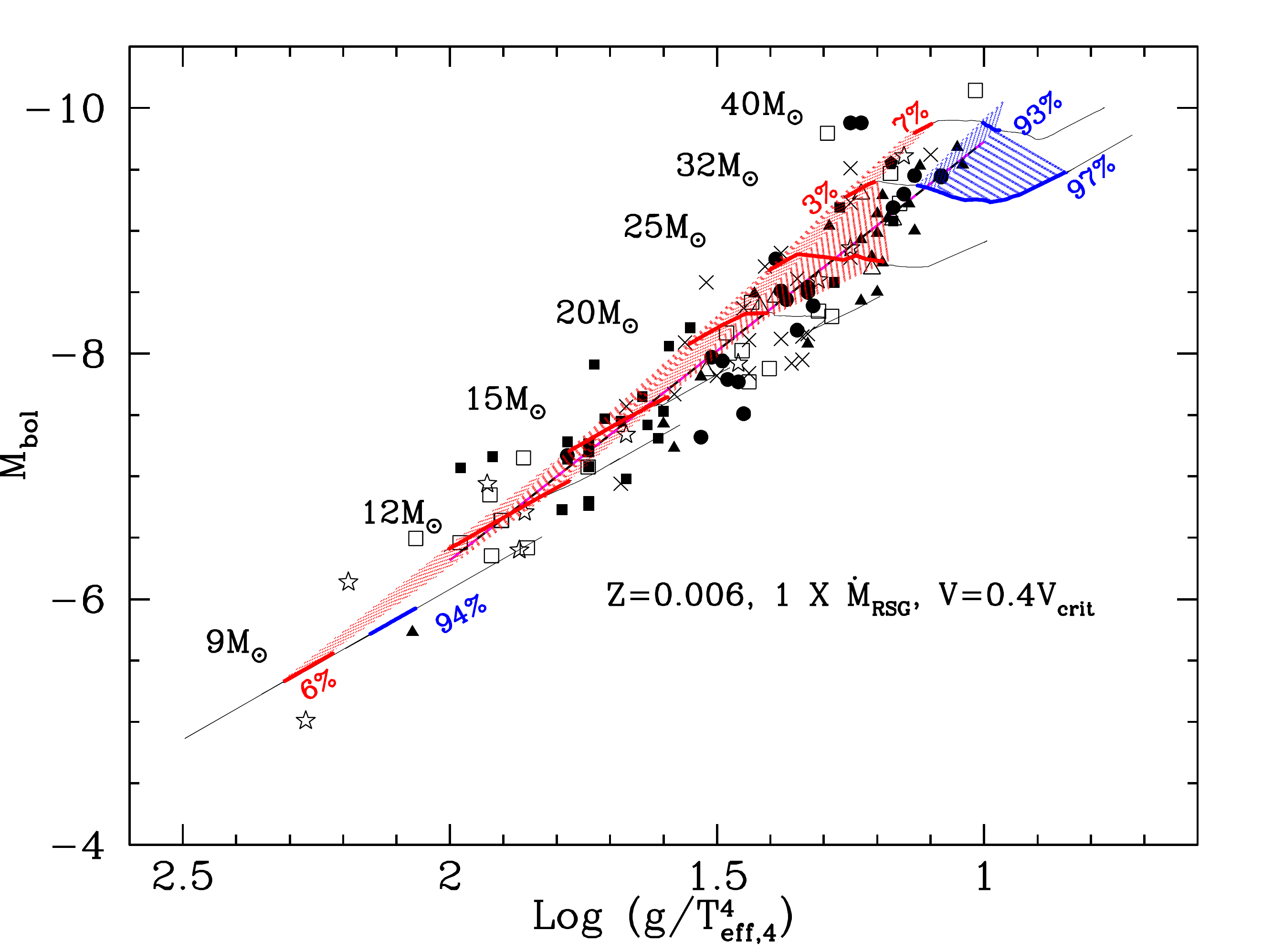}
\includegraphics[width=.38\textwidth, angle=-90]{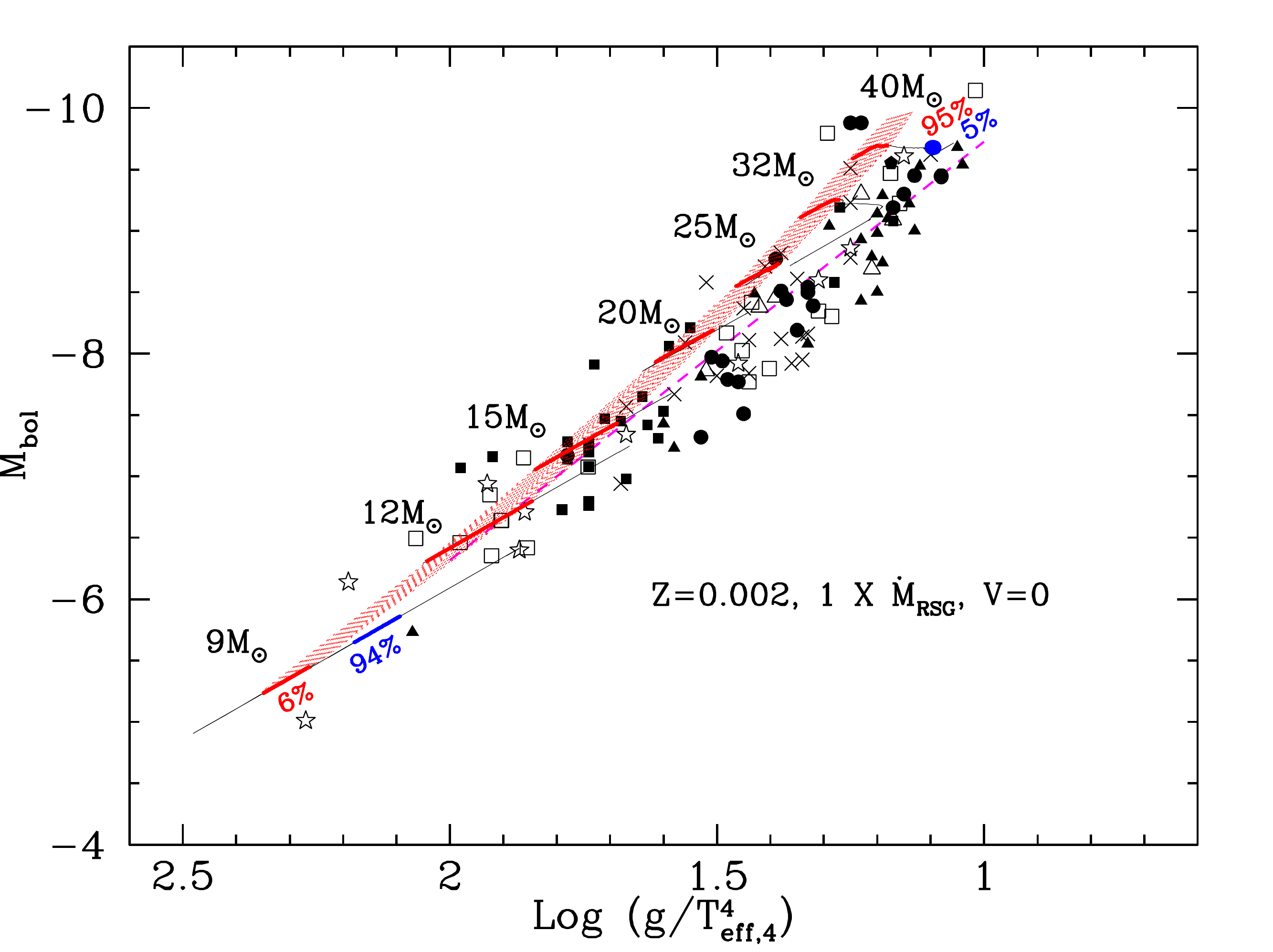}\includegraphics[width=.38\textwidth, angle=-90]{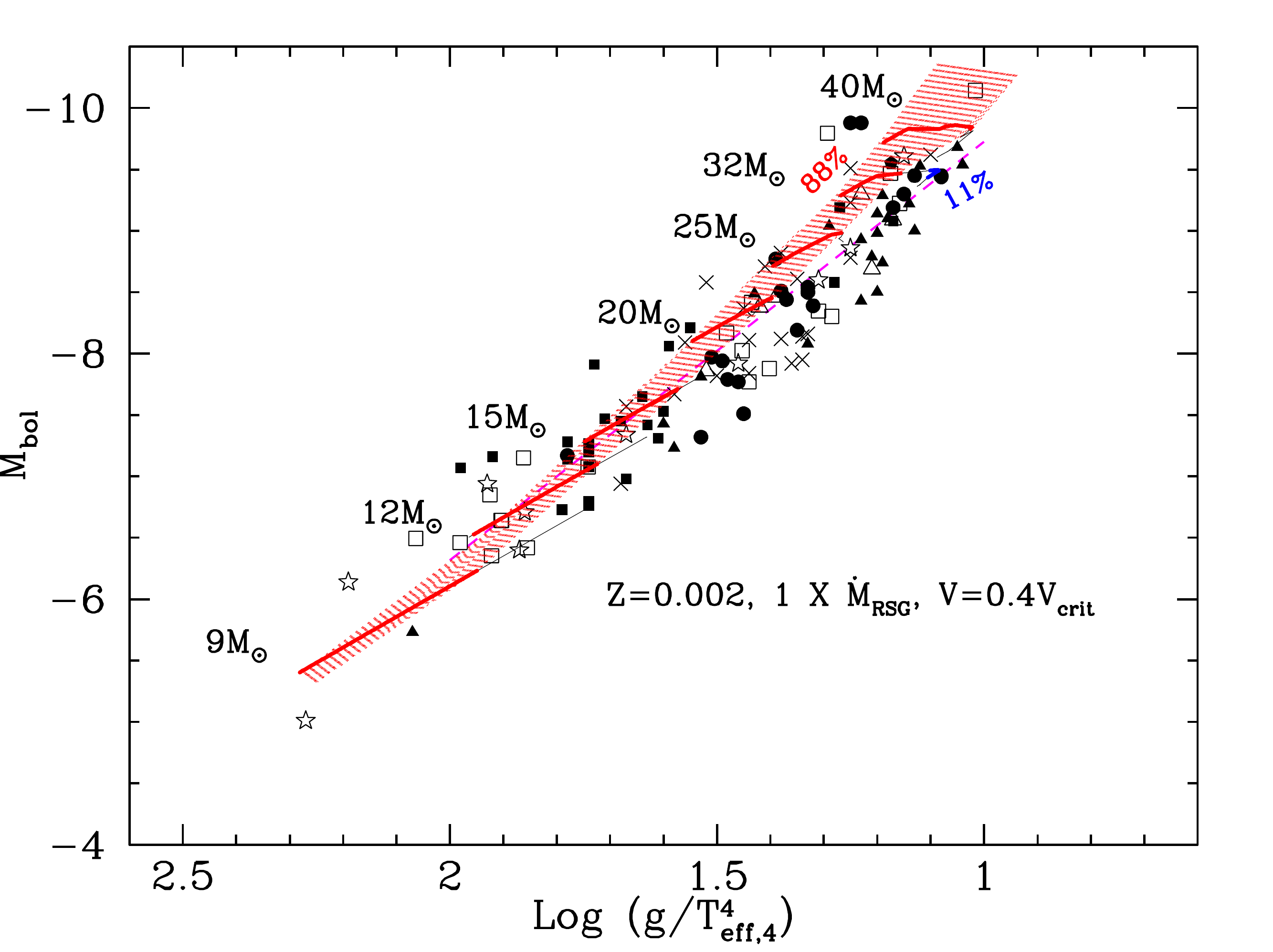}
   \caption{{\it Upper left panel:} Same as Fig.~\ref{fig14m1} for models without rotation, standard RSG mass loss rates but for a metallicity $Z=0.006$. 
   {\it Upper right panel:} Same as left panel but for models with rotation.
   {\it Lower left panel:} Same as Fig.~\ref{fig14m1} for models without rotation, standard RSG mass loss rates and for $Z=0.002$. 
   {\it Lower right panel:} Same as lower left panel but for models with rotation.}
      \label{fig62m1}
\end{figure*}

Recently  \citet{Georgy2012} and \citet{Meynet14} presented tracks with RSG mass-loss enhanced by a factor 10 and 25. 
The rationale for computing such models is the observational fact that  some core collapse supernovae have a yellow or even sometimes a blue progenitor in the luminosity range where
RSG stars are found. A famous example is the progenitor of SN 1987A \citep[see e.g.][]{Walborn89}.
A way to explain the colors of these progenitors is through RSG enhanced mass loss rate models. 
 The mass loss rate enhancement can be due to the presence of a close companion which
will trigger the removal of part of the RSG envelope \citep[see e.g. ][]{Pod1990}, or to some other physical mechanisms, like pulsations
\citep[see e.g.][]{Yoon2010} that may occur in single stars. The enhancement factors in the range from 10 to 25 produce mass loss rates that are compatible with
some spectroscopically estimated mass loss rates of RSGs. While these measurements may not be representative for the whole RSG-phase, they were used as a guideline to explore consequences of
strongly enhanced mass-loss.

It is, thus, tempting to investigate the effects of strongly enhanced RSG mass-loss on the FGLR. The results are shown in Fig.~\ref{figmdot}.
As expected already after the discussion in the previous subsection, the areas corresponding to group 2 are shifted to the right in the FGLR plane and lifetimes during this phase are increased.
(Of course, the areas linked to group 1 BSG remains the same as before in Fig.~\ref{fig14m1}). We can indeed safely conclude that the majority of BSG forming the observed FGLR cannot be the result
of such an evolutionary scenario.
 
While this makes the strongly enhanced mass-loss as the rule in RSG evolution unlikely, the question arises whether this is still a valid scenario to explain 
the core collapse supernovae (CCSNe) progenitors with a yellow or a blue supergiant. We think of three possible solutions:
the first is that the frequency of yellow and blue progenitors for CCSNe is very low (at least at solar metallicity). 
At the moment the statistics of the distribution of the known progenitors among red, yellow and blue supergiants is still poorly constrained and this possibility cannot be ruled out.
Another possibility is that that yellow and blue CCSNe progenitors explode in a very short phase immediately after the RSG phase. It seems that the 
progenitor of the supernova 1987A, which was a red supergiant 20 000 years before exploding as a blue supergiant \citep[see e.g. the discussion in][]{Morris07}, could be an example for such a case.
Finally, a third possibility is a scenario in which less mass is lost during the RSG phase but still the star would evolve back to the blue making
blue supergiants with a moderate decrease of the $M/L$ ratio, still compatible with the small scatter of the flux-weighted gravity-luminosity relation.
We shall come back on that discussion in the conclusions of this paper.

\subsection{Models for SMC and LMC metallicities}\label{SectionSMCLMC}

Flux-weighted gravity-luminosity relations for metallicities lower than solar, obtained from models with and without rotation, are shown in Fig.~\ref{fig62m1}. 
We see that the group 1 BSG, in case of the non-rotating stellar models for both $Z=0.006$ and $0.002$, define again a narrow FGLR sequence
although slightly wider than the one obtained in the solar metallicity case. This comes from the fact, that in these lower metallicity models, a larger fraction of
core He-burning occurs during the BSG group 1 phase \citep[see for instance the discussion in ][]{PVI2001}. This means that the crossing of the HR diagram on the way to the RSG takes a longer time and the lifetime as
a group 1 BSG is much larger. As a result, the stars have more time to lose mass and with the $M/L$ ratio decreasing, the group 1 FGLR becomes wider.

Mass loss by stellar winds is not the only player in this game. Rotational mixing is also a key factor. For instance, it has been shown by \citet{Meynet13LNP} that
depending on the sets of diffusive coefficients used for describing mixing very different situations can appear. In  all the cases except one, rotation
tends to favor a blue location for a significant part of the core He-burning. Only in one case, rotation favors a rapid crossing of the Hertzsprung gap
and a core He-burning phase that occurs mostly during the red supergiant phase and further stages. The present models have been computed with
one of the prescriptions favoring a blue location.  This produces a wider group 1 FGLR. It also produces a bump around
the $25\,M_\odot$ model at $Z=0.006$.  At $Z=0.002$, this bump shifts to higher initial masses.
This behavior can be explained by two facts: first,  all other ingredients being kept the same, 
rotational mixing, which favors a blue location, is more efficient in high mass stars (hence the bump in the high mass range). On the other hand,
mass loss by stellar winds increases with the mass and the metallicity. At $Z=0.006$, above about $30\,M_\odot$, stellar winds overcome the effects
of the rotational mixing (for the initial rotation considered here), reduce the time spent as a group 1 BSG  and thus the bump disappears. At $Z=0.002$, the mass loss rates are smaller, thus the bump
shifts to higher initial masses. 

\begin{figure*}
\centering
\includegraphics[width=.60\textwidth, angle=-90]{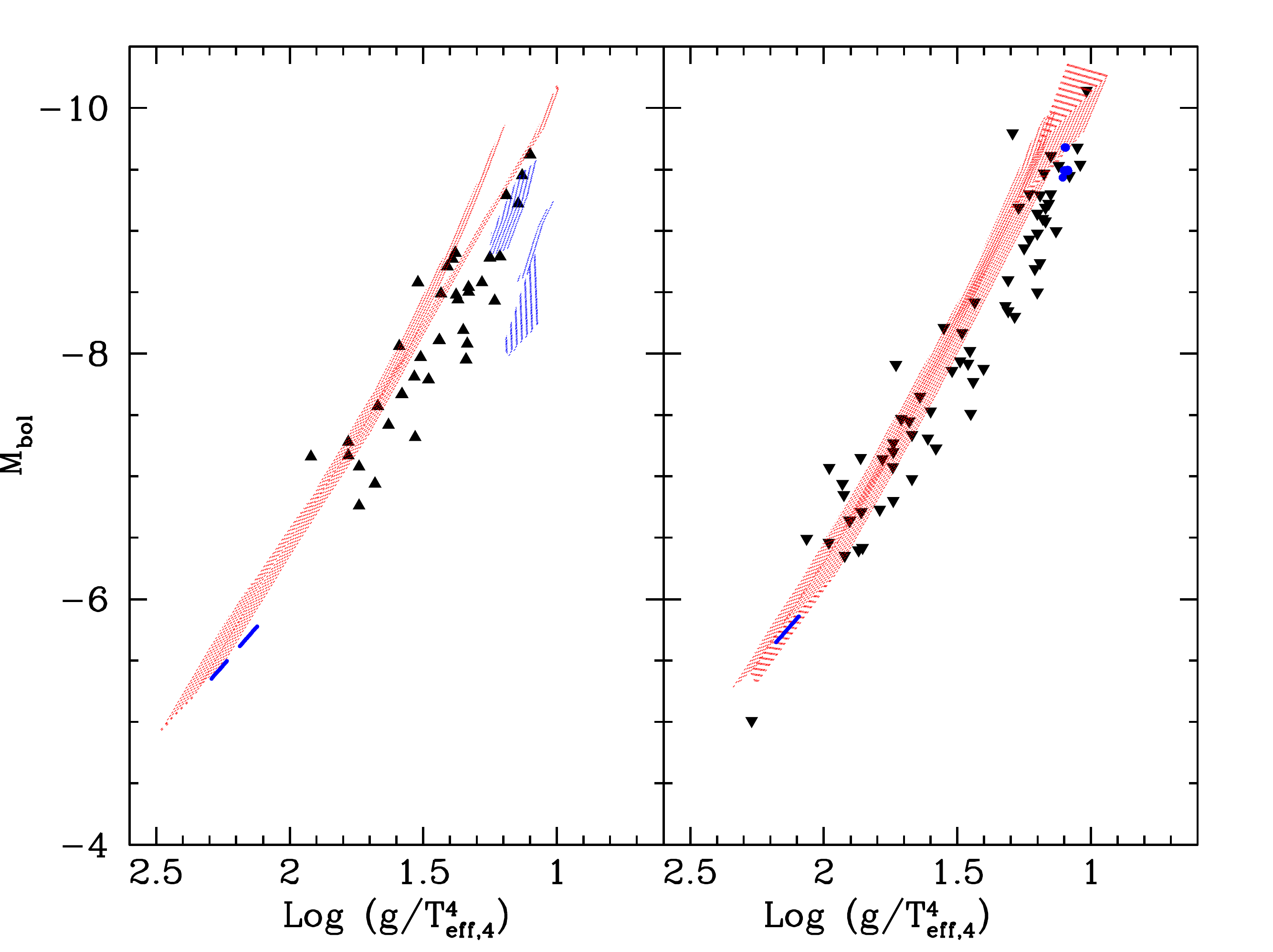} 
   \caption{Distributions of observed blue supergiants in the  $\text{M}_\text{bol}$ versus $\log(g/T_\text{eff,4}^4)$ plane (triangles, see references in Fig.~\ref{lm}). The shaded zones
   indicate where the blue supergiants are predicted according to the stellar models. {\it Left panel:} the observations correspond to stars with a metallicity equal or larger than about half solar,
   the shaded areas show the predictions for the $Z=0.014$ stellar models (rotating and non-rotating with standard mass loss rates.  {\it Right panel:} 
   the observations correspond to stars with a metallicity smaller than about half solar,
   the shaded areas show the predictions for the rotating and non-rotating $Z=0.002$ stellar models with standard mass loss rates. The blue dots shows the very restricted zone
   where group 2 BSGs lay in this diagram, see lower panels of Fig.~\ref{fig62m1}.}
      \label{comp}
\end{figure*}

As can be seen from  Fig.~\ref{fig62m1}, this broadening of the group 1 sequence is compatible with the observed scatter of the flux-weighted gravity-luminosity relation for both metallicities $0.006$ and $0.002$.

Concerning the group 2 BSG, we have only a very modest shift to the right for the positions of these stars in the $\text{M}_\text{bol}$ versus $\log g/T_\text{eff,4}^4$ plane.
This comes from the fact that, due to the effect just discussed above (long duration in the group 1 BSG stage), the stars enter at a late stage of the
core He-burning phase into the RSG phase, so that little time is left for them to lose large amounts of mass. Despite this, the most massive models can still evolve back to the blue
but with $M/L$ ratios which are barely lower than for group 1. This is very different from the solar metallicity case and in better agreement with the observations.

If we compare the group 1 (red color coded)  sequences for the $Z=0.014$,  $Z=0.006$ and $0.002$ models with or without rotation,
we see that they produce very well defined and very similar sequences with a very small scatter.
Therefore, one of the main conclusions of this section is that, the flux-weighted gravity-luminosity relation built with the group 1 blue supergiants
does not depend much on metallicity.

Using the observed blue supergiants FGLR sequence, we can check this point, at least with regard to the observed scatter of the relationship.
The empirical sequence is made of stars with metallicities quite diverse, between one tenth of the solar metallicity and above the solar metallicity.
If we plot in separate figures those stars having a metallicity above half solar and below half solar, we obtain the two distributions shown in Fig.~\ref{comp}.
We see that the solar metallicity objects do not show a larger scatter than the ones at low metallicity.
This indicates that the solar metallicity models produce a too dispersed group 2 blue supergiants especially with the rotating models. As already
indicated above, this may be due to too large mass-loss rates adopted during the red supergiant stage (even in case of the standard mass loss rate case).

\subsection{Stellar evolution models from other groups}

It is important to check whether the conclusions obtained so far are independent of the specific properties of the stellar evolution models used. To this extent we compare
the FGLR obtained from the \citet{Ekstrom2012} with the FGLRs constructed from two sets of independent models, published by \citet{CL2013} and \citet{Brott2011}. This comparison is carried out in
Fig.~7.

There are many differences between the three sets of models. We mention here the most significant ones.
Brott's code uses the Ledoux criterion for convection and includes semi-convection, while the two others use Schwarzschild. 
The models differ by the amount of overshooting considered. In \citet{Ekstrom2012}, the overshooting parameter was chosen equal to 0.1 H$_p$, while in 
\citet{CL2013}, it was chosen equal to 0.2 H$_p$ and in \citet{Brott2011}, equal to 0.335 H$_p$. In addition, there is a difference in the adopted initial metallicity on the ZAMS.
\citet{Ekstrom2012} and \citet{CL2013} have an initial metallicity mass fraction of 0.014, whereas the models by \citet{Brott2011} start with 0.0088. The details of the differences of the models 
including rotation are discussed further below.

The models by \citet{Brott2011} allow to make comparisons only for the group 1 BSG, since the computations of the stellar models were stopped at the beginning of the core He-burning phase.

Despite these significant differences, the three grids of non-rotating stellar models predict very similar positions in ($\log g/T_\text{eff,4}^4$, $\text{M}_\text{bol}$)-plane as can be seen on the 
upper left panel of Fig.~7. The scatter for the group 1 BSGs between the models is well below the observed one.

The group 2 BSGs for the non-rotating models of \citet{CL2013} exists for the 25, 30 and 40 M$_\odot$ models. These BSGs are at much lower values of $\log g/T_\text{eff,4}^4$ than the group 1 BSGs. The time spent in the group 2
corresponds to 61\% of the total BSG lifetime for the 25 M$_\odot$ model and to 90\% for the 30 and 40  M$_\odot$ stellar models.
This is in a remarkable qualitative agreement with the models by \citet{Ekstrom2012}\footnote{Note that the two grids of models do not use the same mass loss rates during the RSG phase, see the precise references in \citet{Ekstrom2012} and \citet{CL2013}.}. 

In the lower left panel of Fig.~7, comparisons between rotating stellar models are carried out. We note that the initial rotation velocities considered by the different authors 
are different. In \citet{Ekstrom2012}, 
the initial velocities span the range between 248 and 314 km s$^{-1}$  for the initial masses between 9 and the 40 M$_\odot$ models (see their table 2). The models by \citet{CL2013} 
have initial velocities on the ZAMS equal to 300 km s$^{-1}$. The models by \citet{Brott2011}, shown on Fig.~7, have initial velocities between 216 and 226 km s$^{-1}$. The physics of rotation
is not the same in these three grids. The models by \citet{Brott2011} use a very efficient transport for the angular momentum driven by a dynamo activity in radiative layers 
following the theoretical approach by \citet{Spruit2002}. These models rotate nearly like solid bodies during 
the Main-Sequence phase. The
models by \citet{Ekstrom2012} and \citet{CL2013}, on the other hand, consider transport by shear turbulence and meridional currents along the theory proposed by \citet{Zahn1992}. 
These models present differential rotation as a function of radius during the whole evolution. 

Again, despite these very significant differences, the models predict very similar positions for the group 1 BSGs (see the lower panel of Fig.~7). This is quite remarkable and underlines the robustness
of the theoretical relation with respect to changes in the stellar models. What is even more remarkable is the good agreement also for the predictions concerning the group 2 stars.  Thus we can conclude
that whatever the grid we would have used, very similar conclusions would have been obtained.

We can investigate whether exploring a larger range of initial velocities could increase the scatter predicted by the stellar models. For this purpose, we use
 the models 
by  \citet{Brott2011} which
have considered a large range of initial velocities over the range of initial masses that is of interest here. In the right panel of Fig.~7, we show superposed to the observed BSGs, the lines 
connecting 
models at similar stages of their evolution having different initial velocities. All group 1 BSGs with initial velocities between 0 and 431 km $s^{-1}$ are predicted to be found between the lines labeled by 0 and 431 km s$^{-1}$.
We see that the region is narrow and has a width much smaller than the observed scatter. Most of the stars will likely begin their evolution inside this very large range of initial velocities. Only, when considering
initially very fast rotators, with velocities between about 431 and 542 km s$^{-1}$, would the scatter cover a significant part of the observed scatter. However, the frequency of 
these very fast rotators is likely very small and there successor will not contribute significantly to the observed sample of BSGs. We conclude that using a larger
 range of initial velocities 
would hardly change the conclusions obtained previously based
on non-rotating and moderately fast rotating models. 

Note that  one may wonder whether for the fast rotating models, we should not account for the scatter produced in the luminosity, effective temperature and gravity due to the darkening of the equatorial regions and the brightening of poles,
the so called von Zeipel effect \citep{vonZ1924}. However, even starting with a high initial surface velocity on the ZAMS, in the BSG-phase, the star will have a surface velocity 
well below the critical value and thus the corrections to that effect will be small.

As a final test, we also compare with the 15 M$_\odot$ stellar model computed by \citet{Georgy2013}, which treats rotation in a similar way as \citet{Ekstrom2012}. While at zero 
rotational differences to the \citet{Brott2011} models in overshooting parameter and metallicity have a slight effect, the models are very similar at medium and high rotational velocity.

In summary, we conclude that differences in the physics of the stellar evolution models used do not affect the conclusions obtained from the comparison with the observations.  

\begin{figure*} 
\centering
\includegraphics[width=.48\textwidth, angle=0]{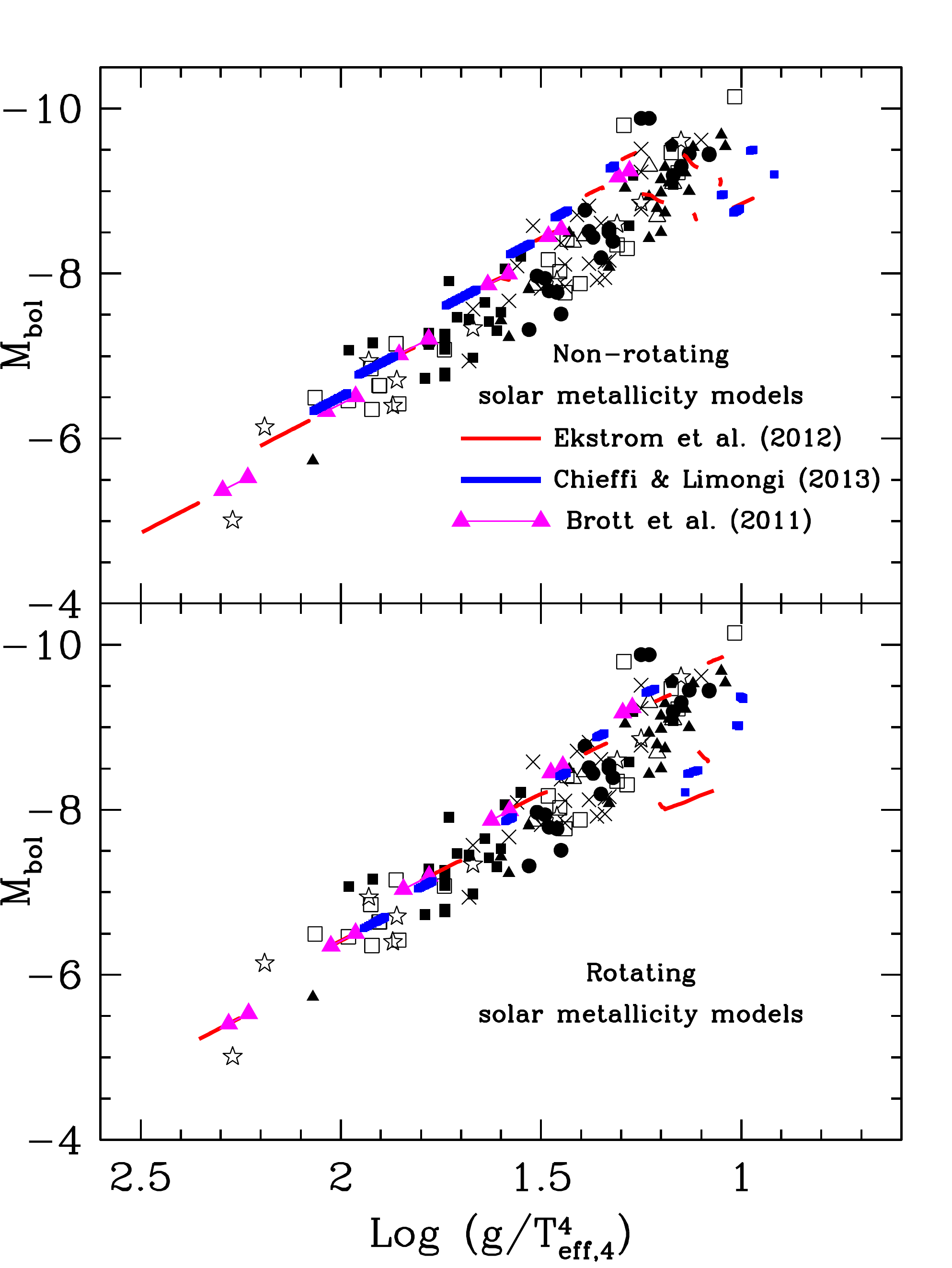} \includegraphics[width=.48\textwidth, angle=0]{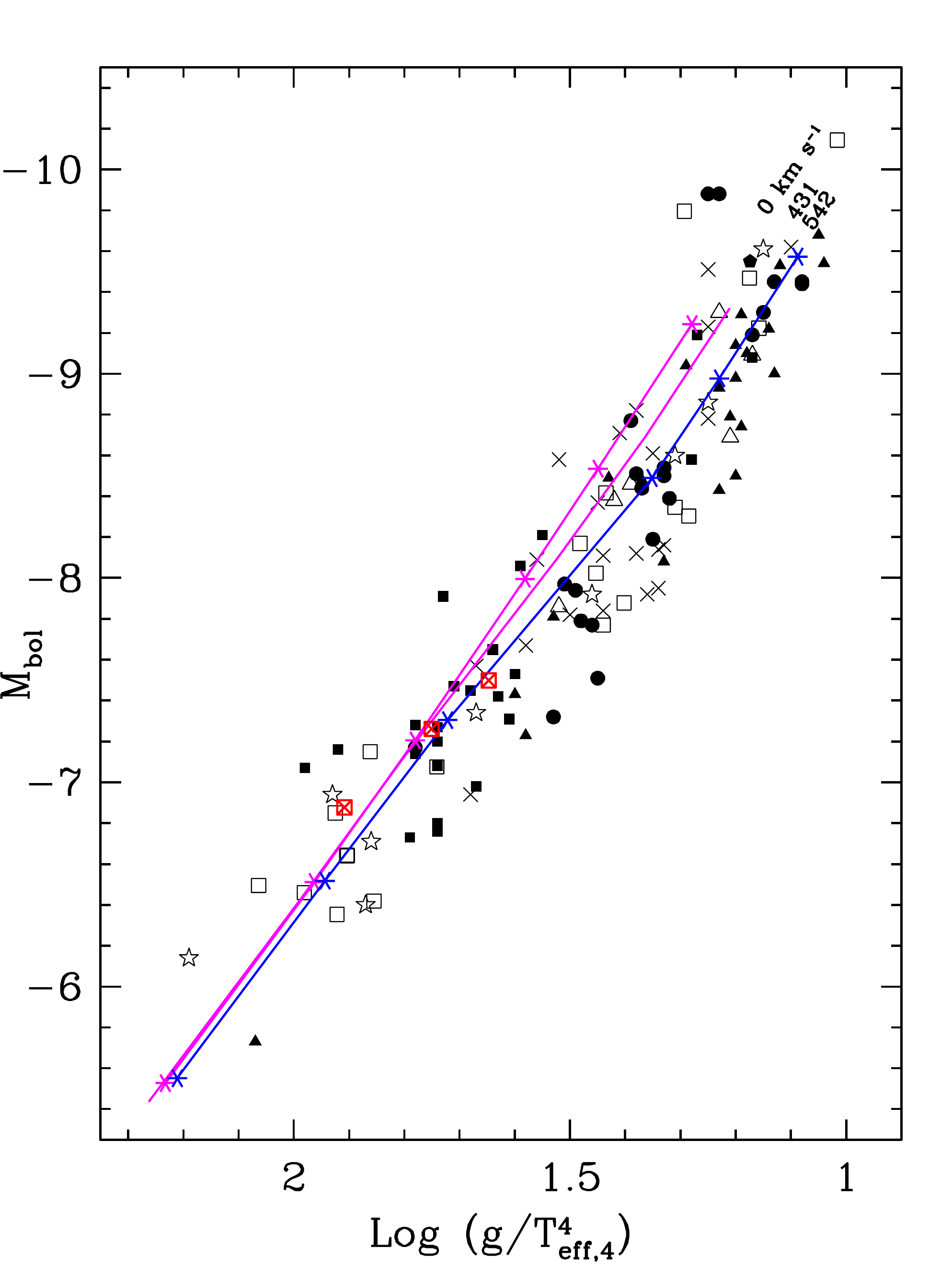}   
 \caption{
 {\it Left panel:} comparisons between the positions of blue supergiants obtained in various grids of stellar models in the ($\log g/T_\text{eff,4}^4$, $\text{M}_\text{bol}$)-plane (see 
text). The stellar grids are those from \citet{Ekstrom2012}, \citet{CL2013}, and \citet{Brott2011}. Initial masses for the \citet{Ekstrom2012} are 9, 12, 15, 20, 25, 32 and 40 M$_\odot$.
For the models by \citet{CL2013} the masses are 13, 15, 20, 30, and 40 M$_\odot$ and for the \citet{Brott2011} models 9, 12, 15, 20, 25, and 35 M$_\odot$. The observations are overplotted as in the 
previous figures. The left upper panel shows evolution models without rotation and the models in the left lower panel include the effects of rotation. The rotating models have 
initial velocities between 216 and 314 km s$^{-1}$ (see text for more details).
 {\it Right panel:} Comparison of the effects of initial rotational velocity for group 1 BSG using the models by \citet{Brott2011}.
The upper (magenta) line labeled by 0 km s$^{-1}$ connects the location in the FGLR-plane predicted by non-rotating stellar models when they enter the BSG domain in the HRD at the hottest effective 
temperature entry point. The middle (magenta) line  labeled by 431 connects the exit points at the coolest effective temperature as group 1 BSG
as predicted by rotating stellar models with an initial velocities between 431 and 475 km s$^{-1}$. The lowest (blue)  line labeled  by 542 is similar to the middle magenta line but for
rotating stellar models with an initial velocities between 542 and 574 km s$^{-1}$ (see text). The asterisk symbols along the upper and lower curve indicate the positions of the 9, 12, 15, 20, 25 and
35 M$_\odot$ stellar models. Added to this plot as red heavy crossed-squares the positions at the same evolutionary stages of 15 M$_\odot$ stellar models with initial velocities equal to 0, 241 and 525 km s$^{-1}$ (from left-bottom to right-up) as
computed by \citet{Georgy2013} with similar physics as in \citet{Ekstrom2012}.
}
      \label{figbrott}
\end{figure*}

\section{Population synthesis models}

In this last section, we use a different tool to compare stellar evolution models with the observations. Up to this point we
have directly transformed stellar tracks into the FGLR-plane and then compared their locations with the observed BSG distribution, while simultaneously considering
stellar lifetimes in the group 1 and 2 BSG evolutionary phases.

Now we advance this comparison by computing virtual stellar populations as they would appear from continuous star formation during the last
70 million years, assuming the initial stellar masses are distributed according to an initial mass function (IMF). For our sample of observed BSG the hypothesis of a constant star 
formation rate during the last 70 million years appears to be reasonable. We also note that we do not intend to compare the relative predicted number of stars at low, middle and high luminosities, as the observed sample is heavily biased by selection effects with regard to luminosity (see below). Instead, our intention is to compare whether the predicted location in the FGLR-plane described by isocontours of high probability at each luminosity agrees with the location of the observed sample.

Such population synthesis models have the qualities (and defects) of the
stellar tracks, however they add two features that otherwise are difficult to assess: 1) they provide quantitative information about the expected number distribution of BSG in the FGLR-plane; 
2) the population synthesis technique easily allows to include the effects of observational errors. Both together will allow us much better to investigate whether models agree with the 
observations or not. We will investigate the cases of $Z=0.014$ (solar metallicity) and $Z=0.002$ (SMC metallicity), with and without rotation assuming standard 
mass-loss rates in the RSG phase.

\begin{figure*}
%: fig TcrcZ014_Mcc_rot.pdf
\centering
\includegraphics[width=.48\textwidth, angle=0]{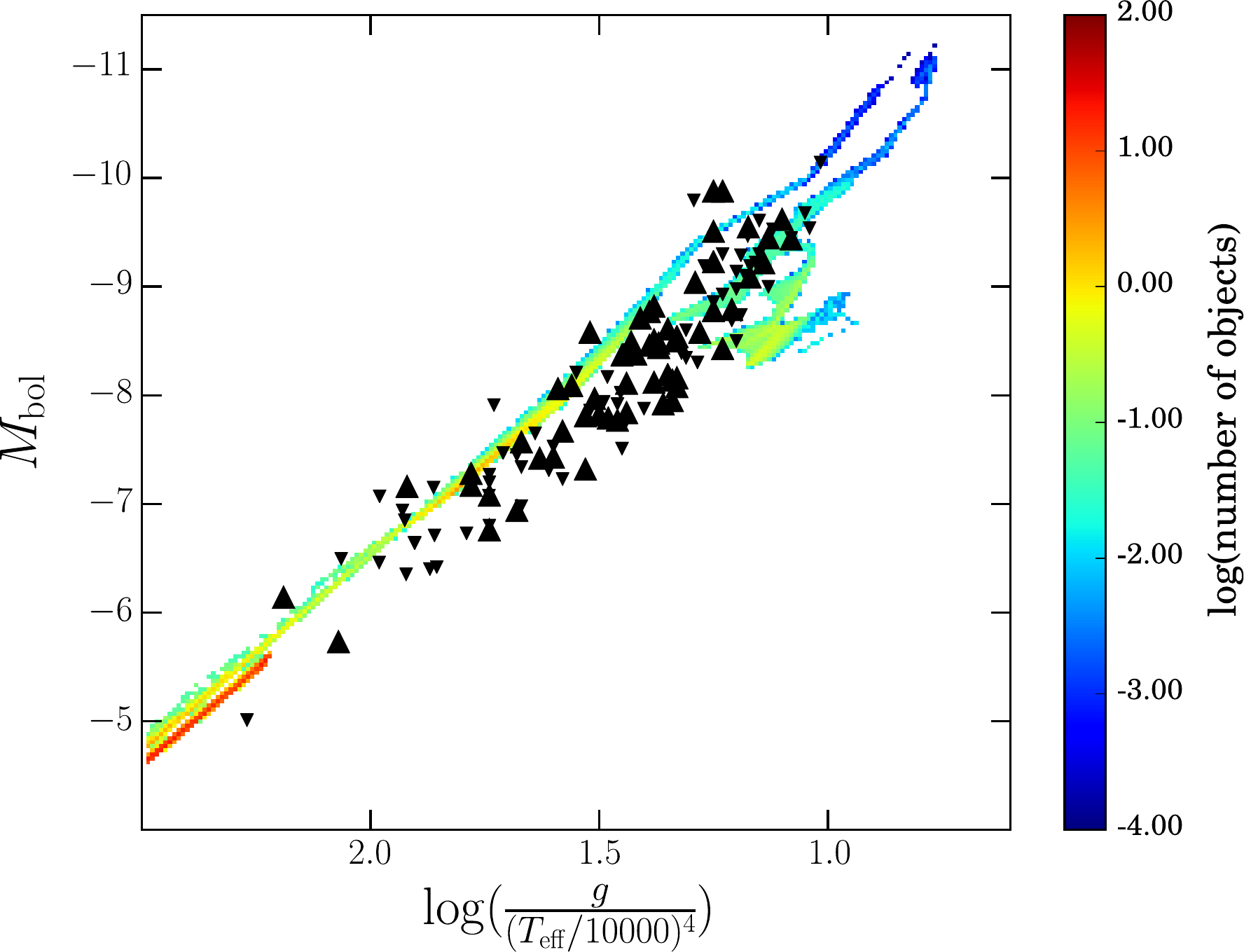} \includegraphics[width=.48\textwidth, angle=0]{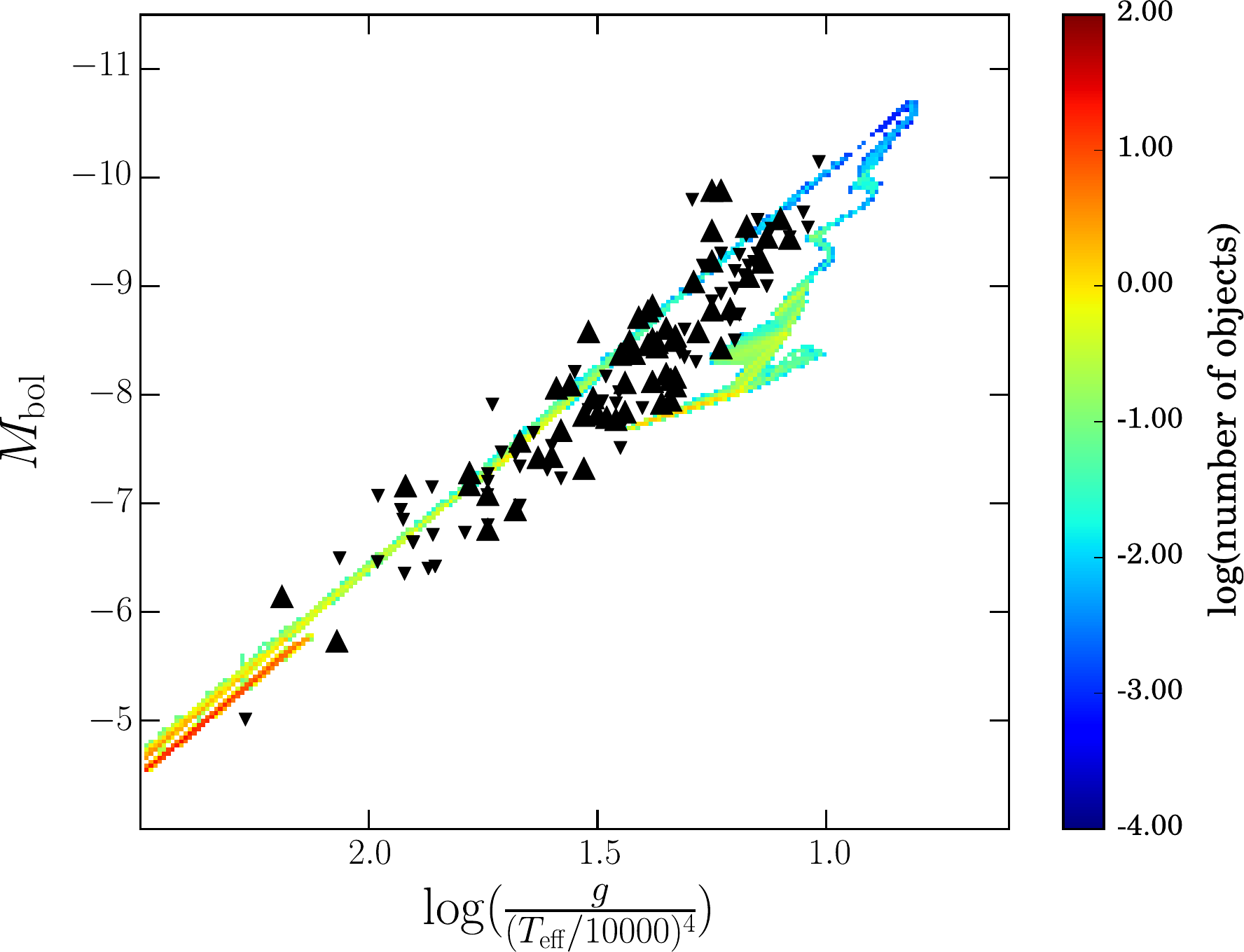}
\includegraphics[width=.48\textwidth, angle=0]{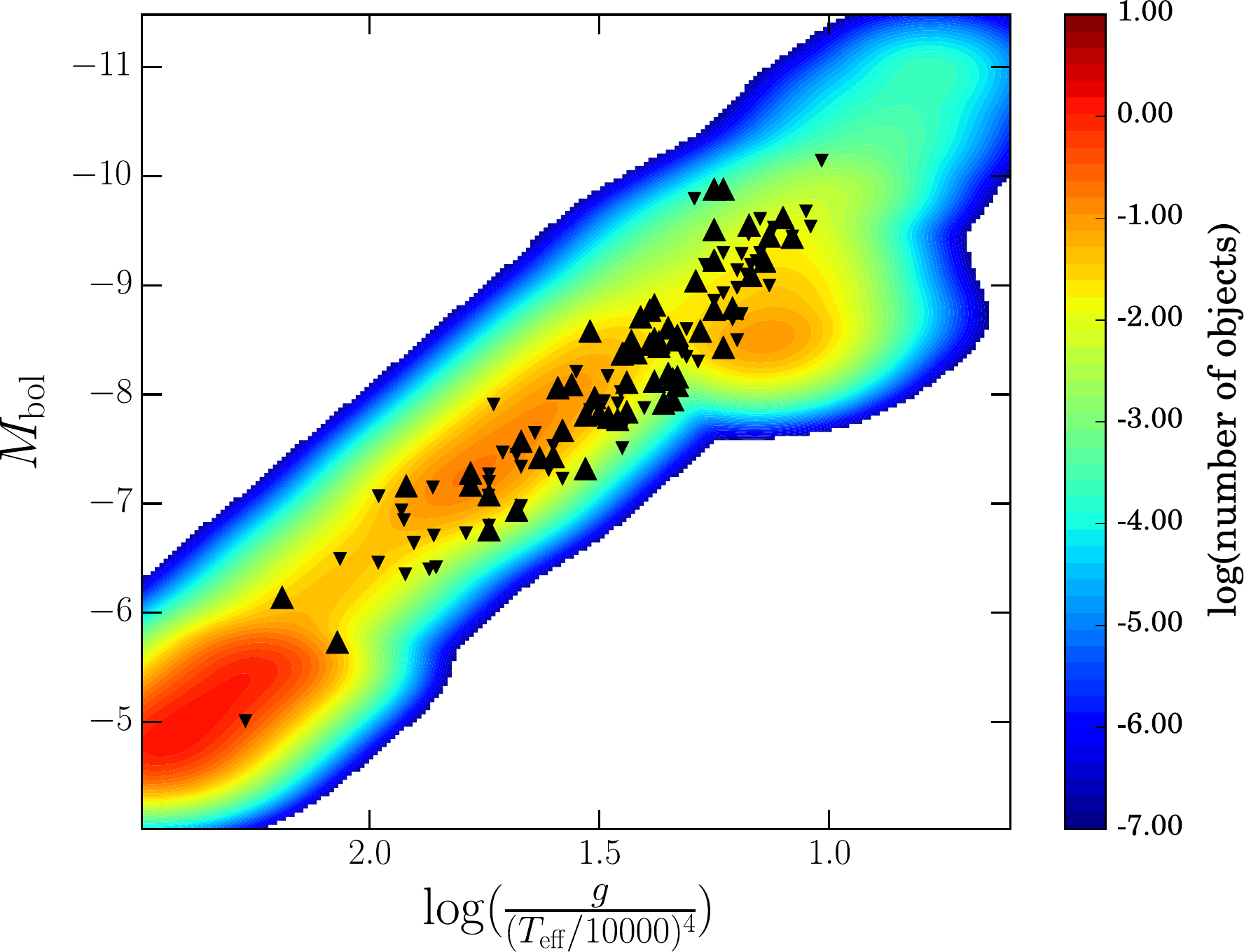} \includegraphics[width=.48\textwidth, angle=0]{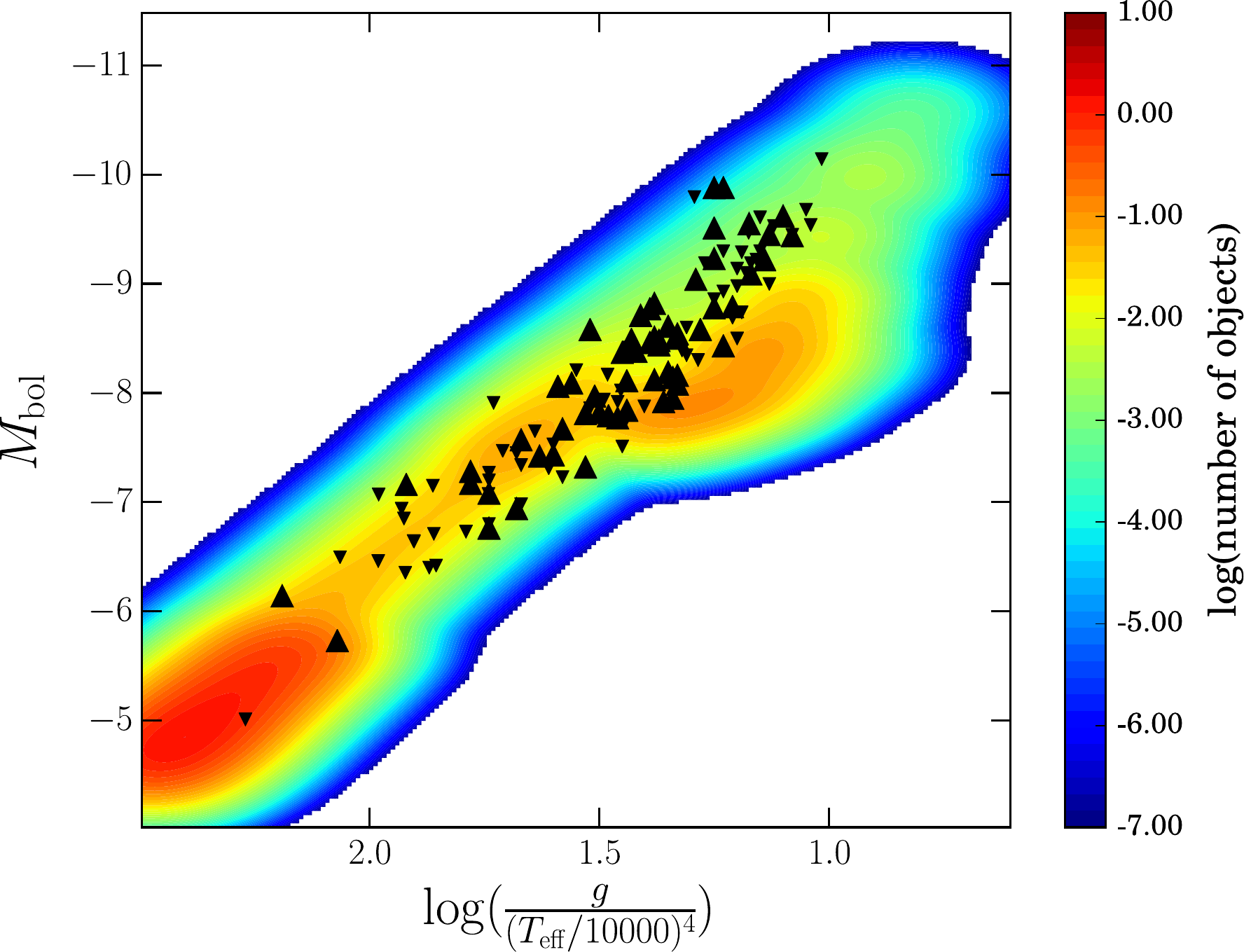}
   \caption{Population synthesis of blue supergiants in the FGLR-plane compared with the observations. The color coded areas indicate the expected relative numbers of BSG.
   Triangles up are "high metallicity" (more than half solar) observed blue supergiants and triangles down are "low metallicity" (less than half solar) ones.
    The triangles corresponding to the metallicity of the plot  are larger. In the present plot for example, the triangles up are bigger.
   {\it Left upper panel:} non-rotating models at $Z=0.014$ calculated with standard mass-loss rates.
    {\it Right upper panel:} Same as the left upper panel but for models including stellar rotation.
    {\it Left lower panel:} Same as the upper left panel but additionally including the effects of observational uncertainties.
    {\it Right lower panel:} Same as the upper right panel but additionally including the effects of observational uncertainties.     
    }
      \label{figdensity}
\end{figure*}

\begin{figure*}
\centering
\includegraphics[width=.48\textwidth, angle=0]{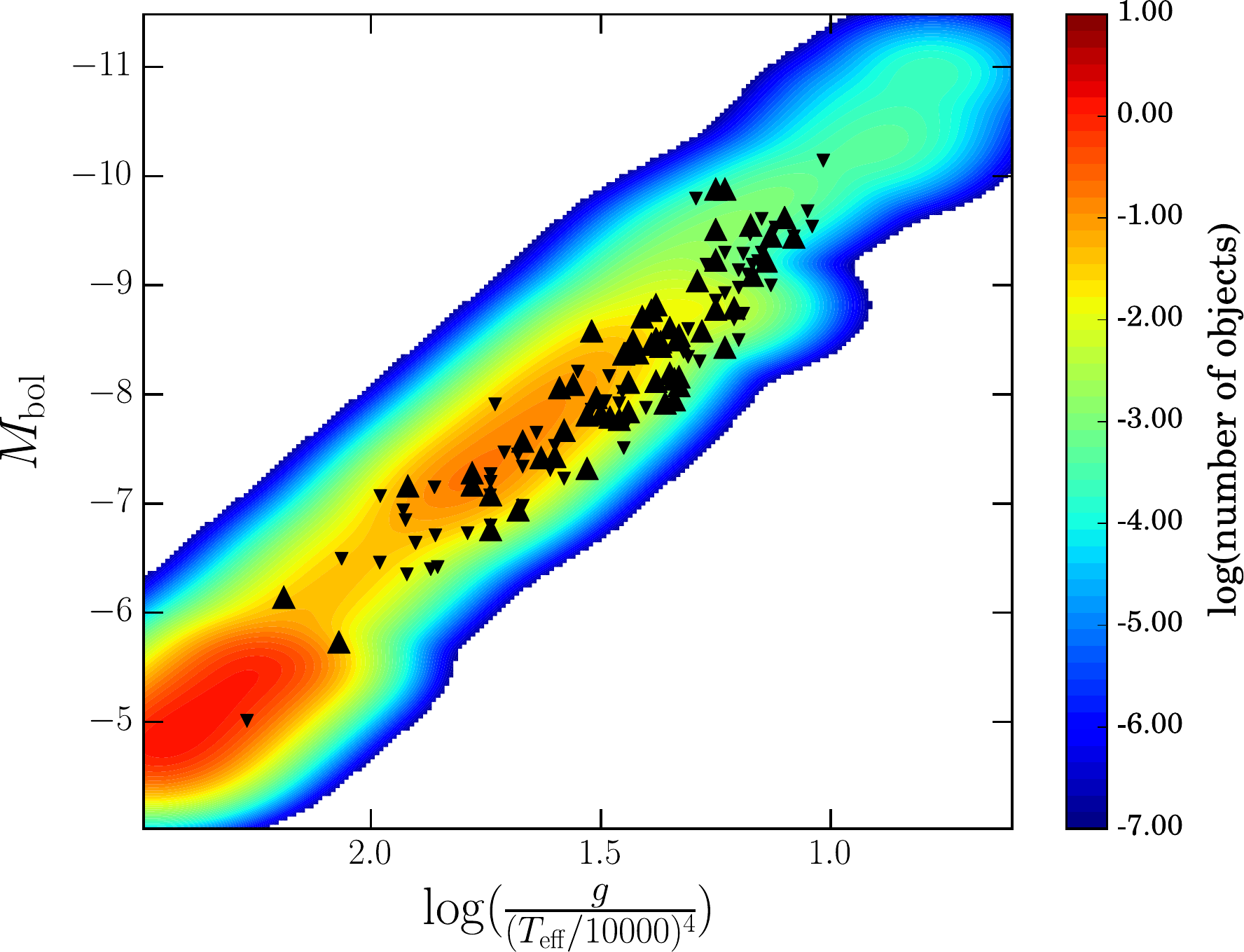} \includegraphics[width=.48\textwidth, angle=0]{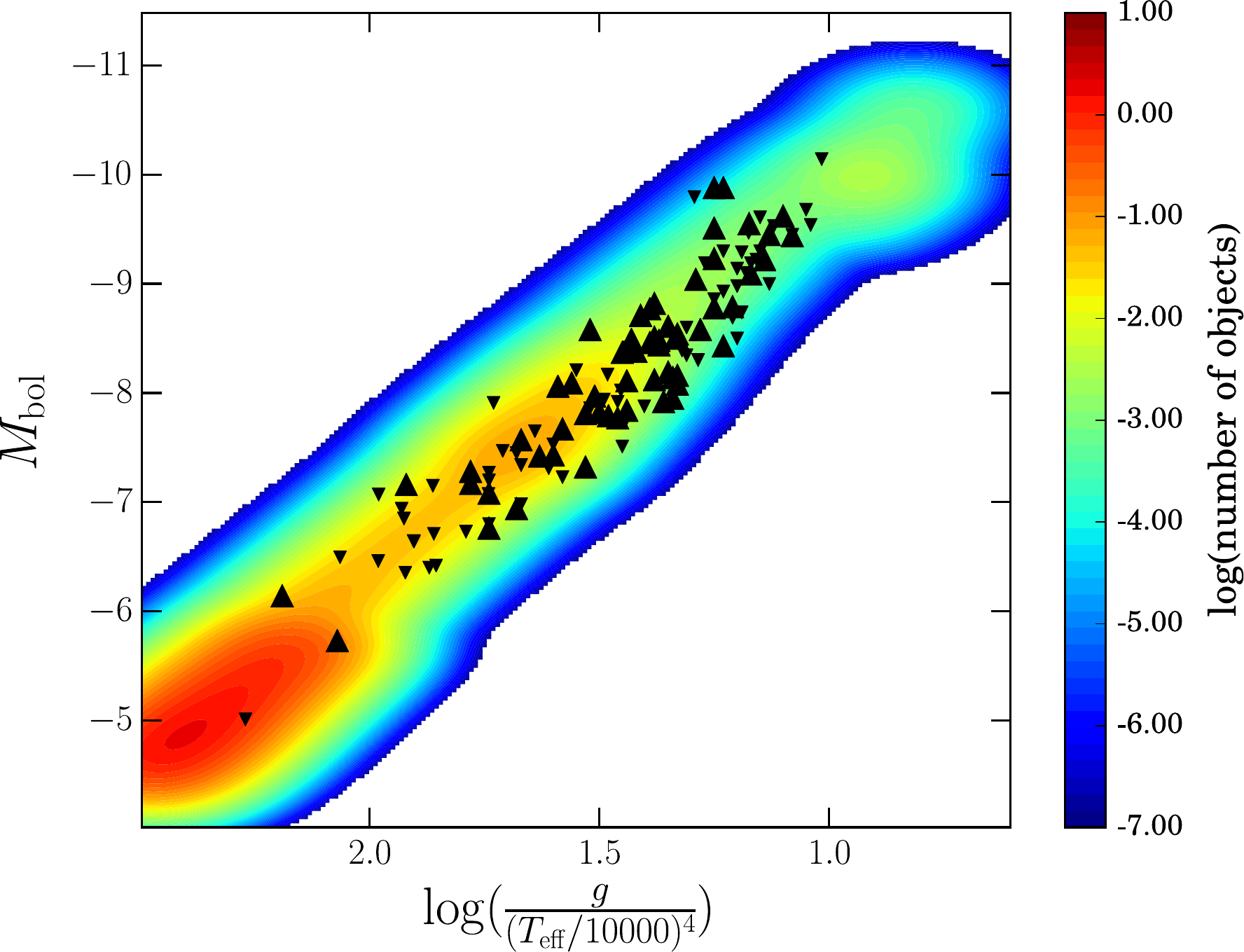}
   \caption{Same as lower panel of Fig.~\ref{figdensity} but assuming that there is no group 2 BSG (see text).
   {\it Left panel:} Tracks without rotation.
   {\it Right panel:} Tracks with rotation.
   }
      \label{essai1}
\end{figure*}

For this purpose, we use the population synthesis tool SYCLIST \citep{syclist14}. Assuming a Salpeter IMF with exponent 2.35 in a mass range from $5$ to $120\,M_\odot$ we divide the main sequence into 10000 mass intervals and follow the entire evolution for each of these masses as function of time using a grid of 
5000 time steps. The FGLR-plane displayed in the subsequent population synthesis plots is divided into 200X200 pixels of $0.0325\,\text{mag}$ in $\Delta\text{M}_\text{bol}$ and $0.01\,\text{dex}$ in $\Delta\log {g/T_\text{eff,4}}$. At each time step the fraction of the 10000 evolutionary tracks with different initial masses falling into one of these 
pixels is noted providing the probability distribution function in the FGLR-plane at each time step by taking into account the original probability on the main-sequence for such tracks to occur. 
Adding up the probabilities of the 5000 time steps we then obtain the distribution function in the FGLR-plane for continuous star formation, which for convenience is normalized to a total number of
1000 stars. Random observational uncertainties are accounted for by spreading the probability value of each pixel by a bi-Gaussian distribution with standard deviations of $0.15\,\text{mag}$ in 
absolute magnitude and $0.075\,\text{dex}$ in flux-weighted gravity, respectively. These values represent average uncertainties of the observed BSG sample (see references given in the caption of Fig.~\ref{lm}). In this procedure, we assume that a star is a BSG if it fulfills the following criteria: 1) it has left the main sequence stage of core hydrogen burning, 2) its $\log T_\text{eff}$ is between 3.9 and 4.4, 3) it is not an extremely helium enriched Wolf-Rayet-like star, i.e. its surface hydrogen mass fraction $X_\text{surf}$ is above 0.3 \citep[same criterion as in][]{Georgy2009}.

Fig.~\ref{figdensity} summarizes the calculation at solar metallicity. The upper panels show the population synthesis results without accounting for observational uncertainties to
provide the information solely coming from the evolutionary tracks, whereas the lower panels include the observational errors and allow for a more realistic comparison. We start the discussion with
the non-rotating models in the upper left panel. The shape of the blue supergiant region is now more complex than in Fig.~\ref{fig14m1}, because of the much finer discretization in stellar mass.
Nevertheless, the overall shape is of course following the one obtained in Fig.~\ref{fig14m1}. The peaks of high densities (yellow and red colors) correspond to stages which are favored either 
because the mass is small (IMF effect) and/or because their duration is large enough (typically this is the case
for the yellow region in-between $\log g/T_\text{eff,4}=1.9$ and $1.5$). At this point it is important to note that the distribution of the observed BSG along the FGLR as a function of bolometric 
magnitude is heavily biased by a selection towards creating a flat distribution in the higher magnitude range for the purpose of distance determinations. This explains why there are only a few 
observed BSG in the magnitude range between -4 to -6 which has the highest probability. These objects would simply require too much spectroscopic observing time.

For both cases, evolution models with and without rotation, the sequence of blue supergiants divides into different channels at higher luminosity corresponding to group 1 and group 2 in 
our previous discussion. (In our population synthesis model, we define a group 2 BSG as a star which had an effective temperature $\log T_\text{eff} < 3.8$ at some point in its past history). The group 2 channels, which are shifted to lower gravities at a given $\text{M}_\text{bol}$, are favored in terms of the expected number of stars as a consequence of the predicted 
stellar lifetimes. We also see that the group 1 BSG sequence for the rotating models is slightly shifted to the right with respect to the non-rotating ones. As already explained above, 
this comes from the fact that models with rotation are overluminous for a given initial mass with respect to non-rotating models 
as a result of rotational mixing. That implies a decrease of the $M/L$ ratio and a shift to the right. 

By including the effects of observational errors the probability of finding BSG in each pixel of the FGLR-plane is spread out over a larger surface. This is shown
in the lower panels of Fig.~\ref{figdensity}. At the first glance, the agreement between the observed and predicted distributions is reasonable. However, with more careful inspection we identify
interesting features common to tracks with rotation and without, but also significant differences.
First, there exists a small valley between the red region in the lower left corner and the diagonal extending upwards towards
the center of the figure. This small valley corresponds to the positions of $12\,M_\odot$ models.
These models evolve very rapidly to the red after the MS phase spending little time of their core He-burning lifetime in the blue.
They also do not come back from the blue and, thus, spend only a modest amount of time as BSG. 
For higher initial masses
a larger portion of the core He-burning phase happens at high effective temperatures and, thus, the lifetime during the BSG group 1 phase increases.
The presence of this valley is likely dependent on physical ingredients of the models as the metallicity, the mass loss rates and the internal mixing.
The present set of observational data does not allow to test this prediction. A more complete spectroscopic survey at lower bolometric magnitudes would be needed for this purpose. 
However, the observed distribution can be used to test other features predicted by the models. For instance, the diagonal probability ridge line in the magnitude range
$-6.5\,\text{mag} \leq \text{M}_\text{bol} \leq -9.5$ should coincide with the location of the observed sequence. Obviously, this is not the case for the non-rotating tracks which predict too bright magnitudes
for $1.9 \geq \log g/T_\text{eff,4} \geq 1.3$ and too faint magnitudes at lower flux-weighted gravities. Fig.~\ref{essai1}, in which we make the ad-hoc assumption that the group 2 BSG does not exist,
demonstrates that the orange high probability area of the models at 
$\log g/T_\text{eff,4} \geq 1.3$ corresponds to group 1 BSG whereas the orange area at lower gravity is created by group 2.

The situation is different for evolutionary models with rotation. Fig.~\ref{figdensity} and \ref{essai1} reveal that the group 1 BSG predicted location in the FGLR-plane agrees much better with the
observations. On the other hand, the prediction for the group 2 BSG is in clear disagreement with the observations. We conclude that, at solar metallicity, models with rotation are in reasonable 
agreement with the observations provided there is a mechanism which suppresses the significant decrease in flux-weighted gravity for high mass group 2 BSG or suppresses the evolution back from the RSG
stage and, thus the existence of group 2, at all.

\begin{figure*}
\centering
\includegraphics[width=.48\textwidth, angle=0]{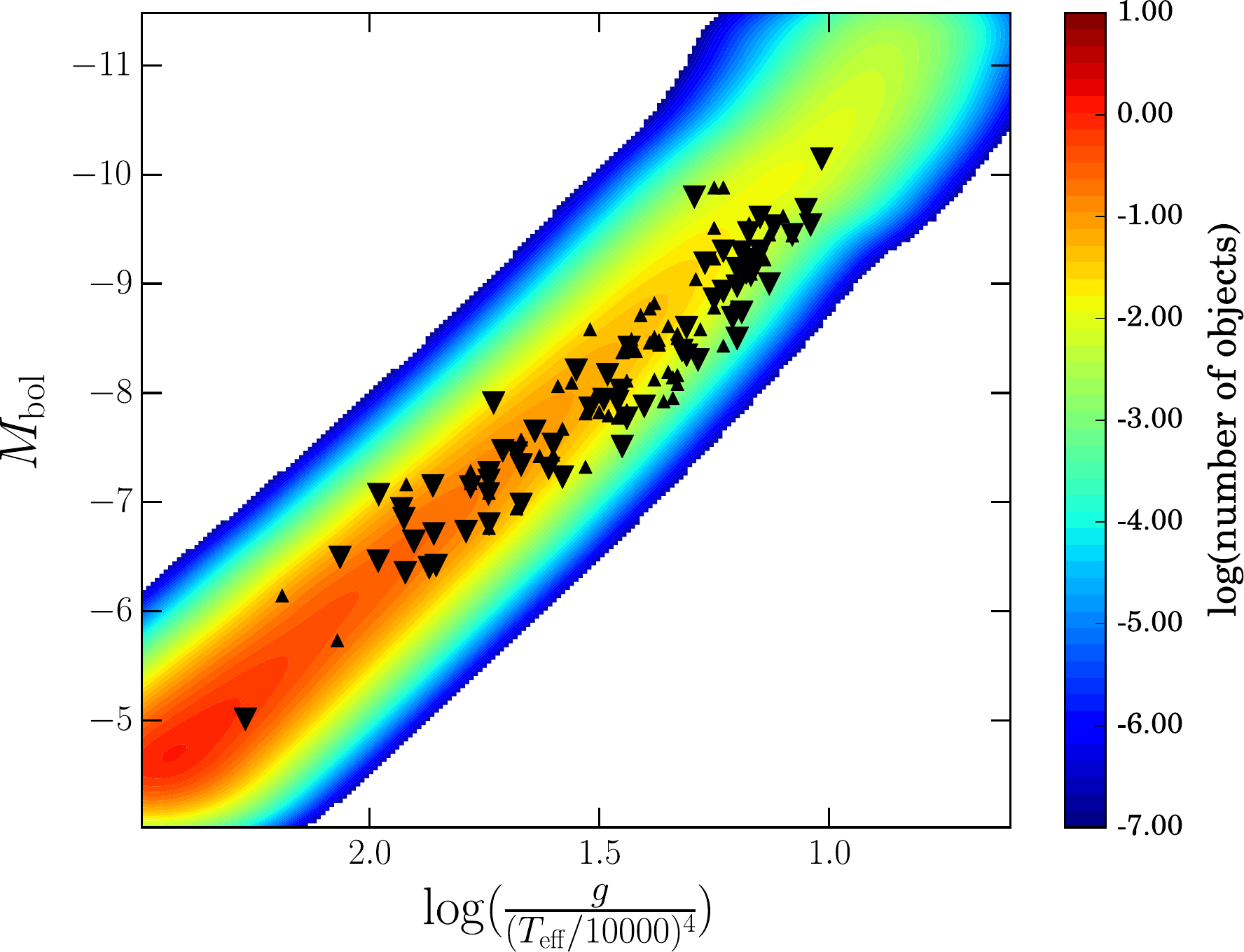} \includegraphics[width=.48\textwidth, angle=0]{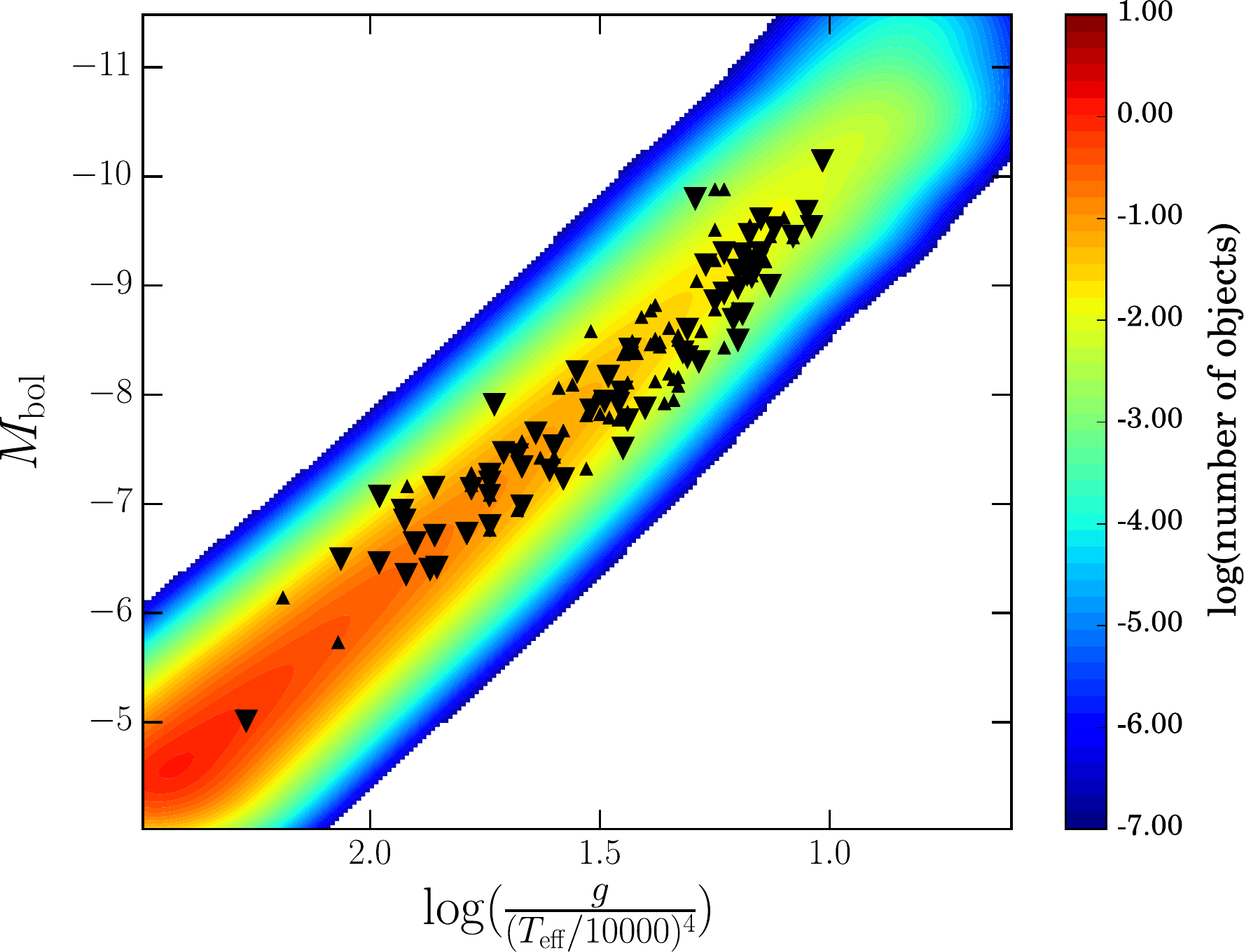}
   \caption{Same as lower panel of Fig.~\ref{figdensity} but for a metallicity of Z=0.002.
   {\it Left panel:} Tracks without rotation.
   {\it Right panel:} Tracks including rotation.
   }
      \label{smc}
\end{figure*}

\begin{figure*}
\centering
\includegraphics[width=.48\textwidth, angle=0]{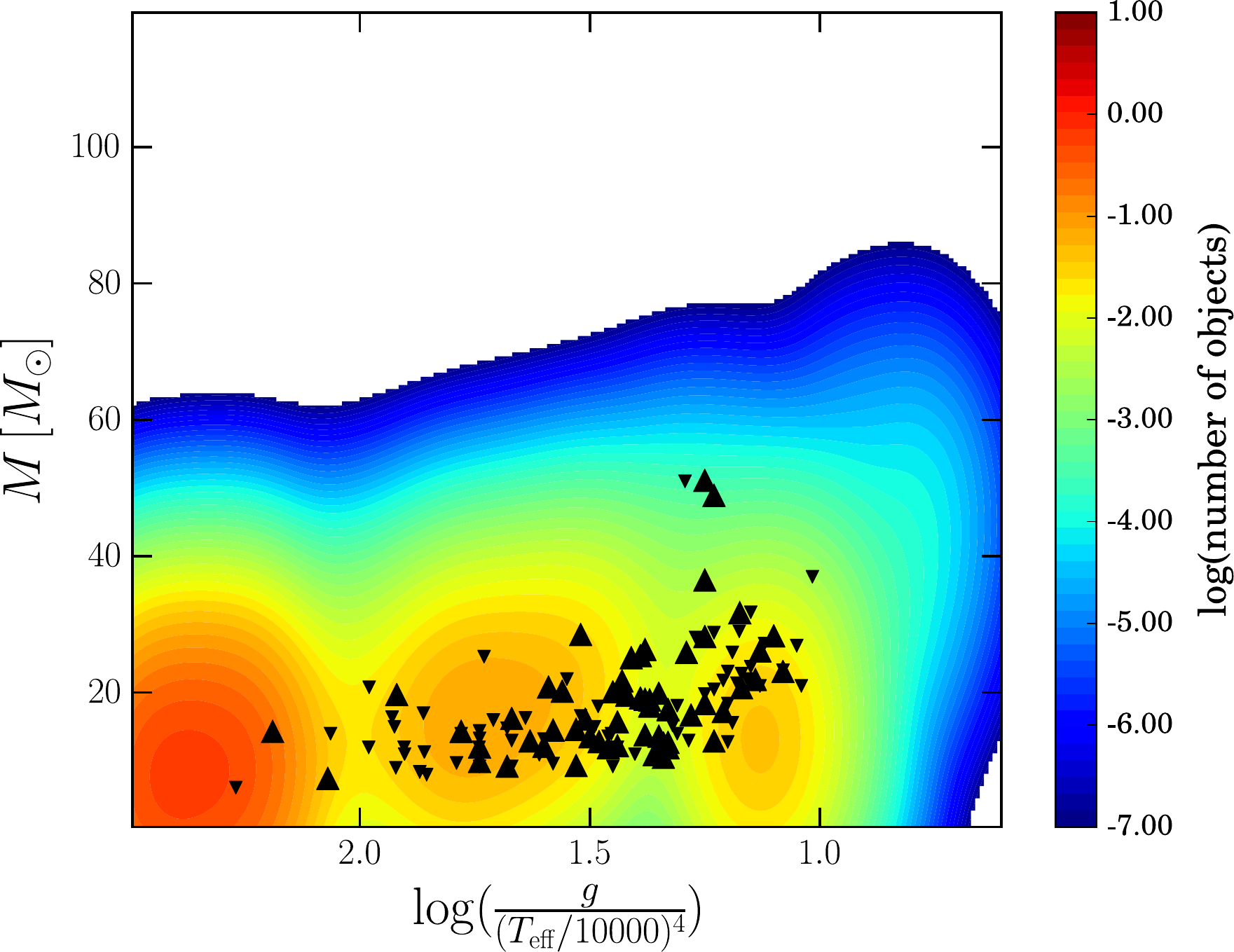} \includegraphics[width=.48\textwidth, angle=0]{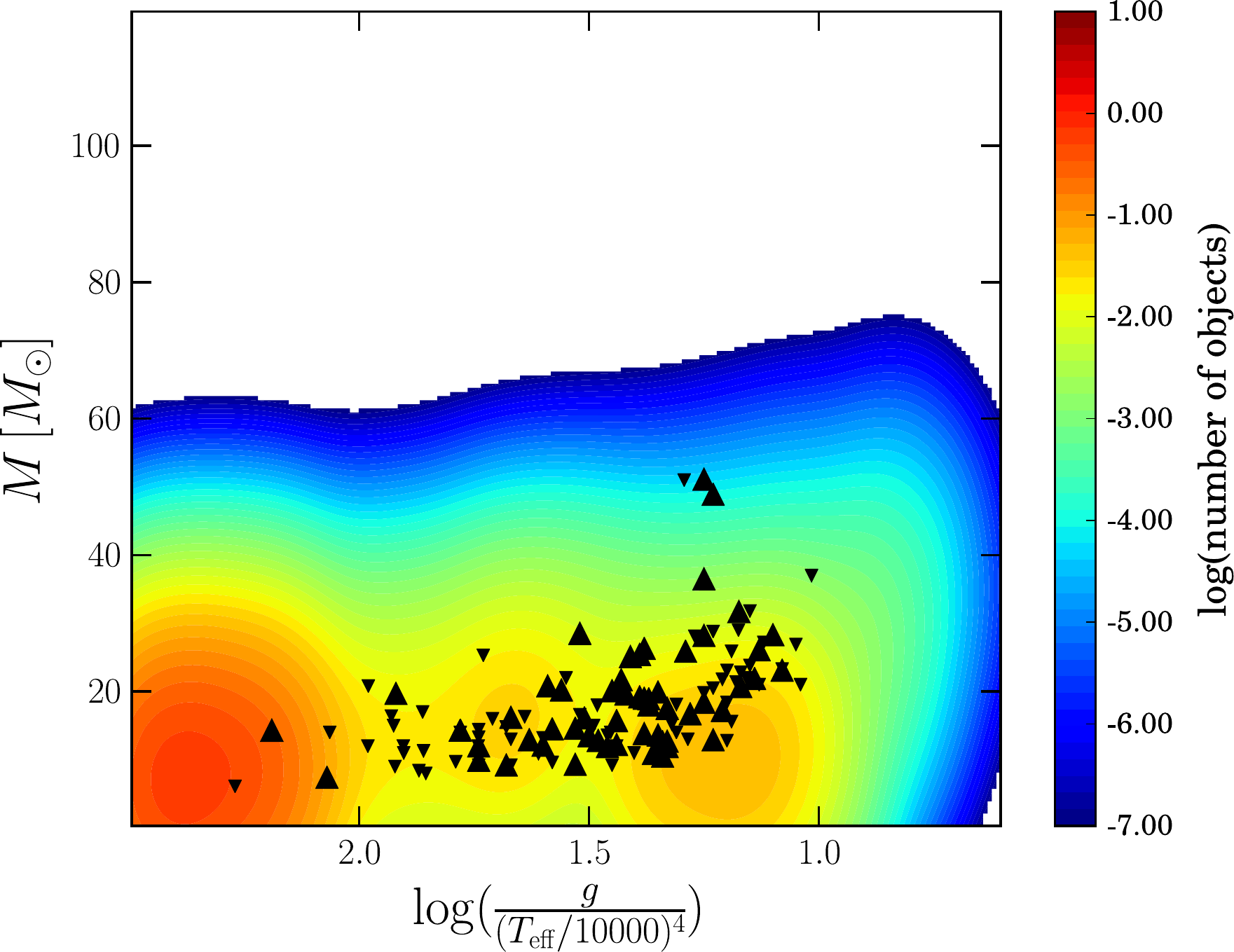}
\includegraphics[width=.48\textwidth, angle=0]{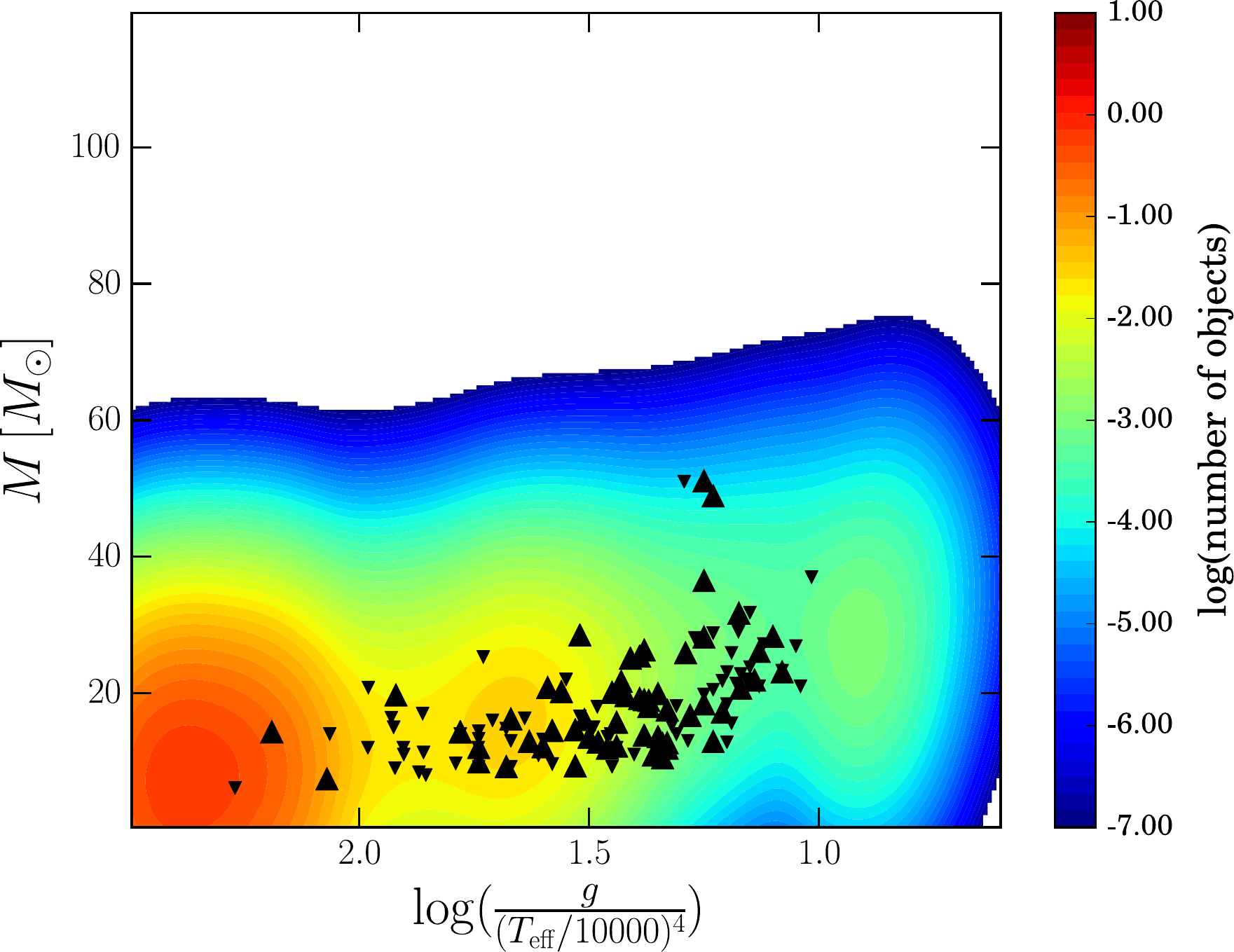}\includegraphics[width=.48\textwidth, angle=0]{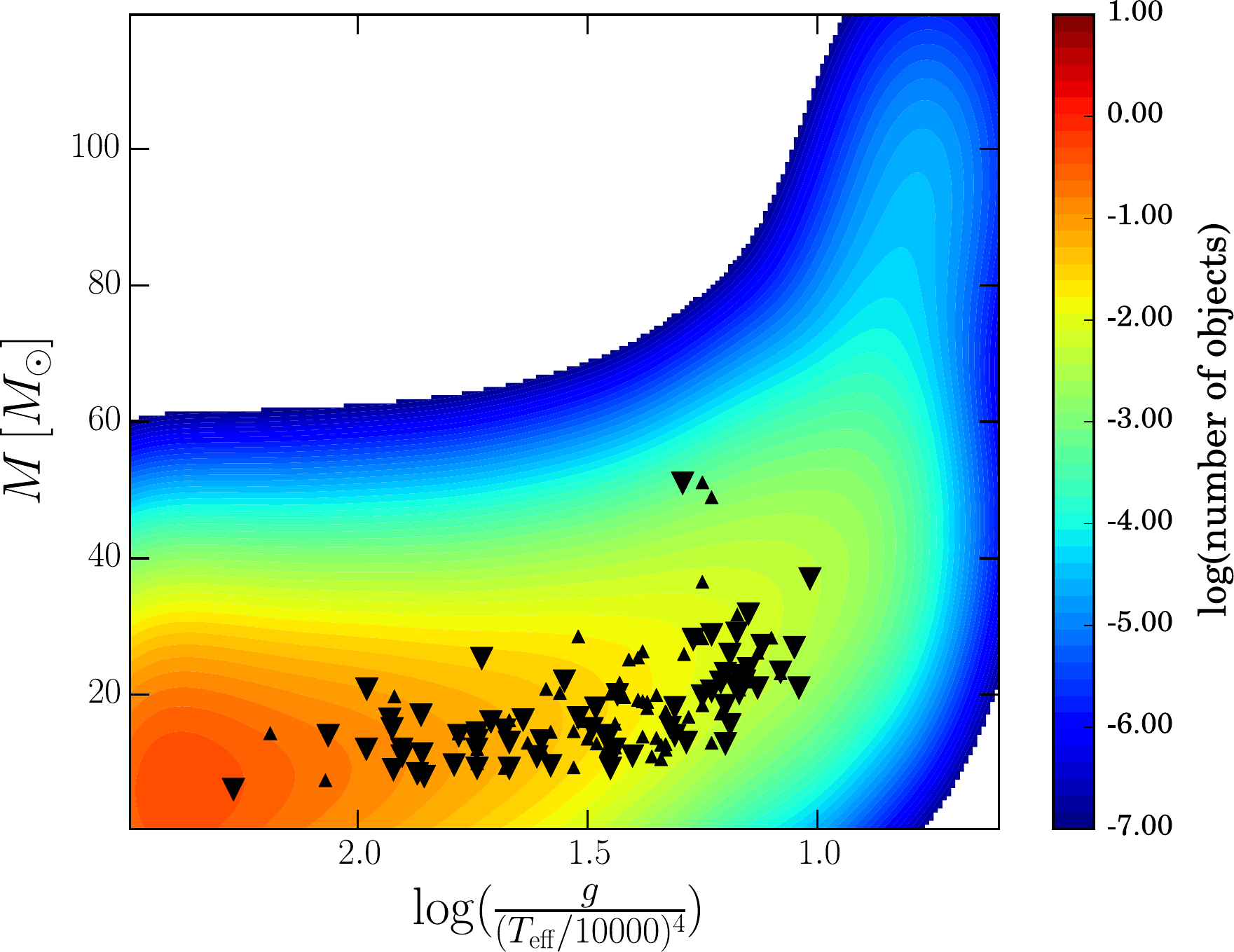}
   \caption{Predicted density plots of evolutionary masses compared with observed BSG spectroscopic masses.
   {\it Upper left panel:} Tracks without rotation at $Z=0.014$.
    {\it Upper right panel:} Tracks with rotation at $Z=0.014$.
    {\it Lower left panel:} Tracks with rotation at $Z=0.014$ without group 2 BSG.
   {\it Lower right panel:} Tracks with rotation at $Z=0.002$.
   }
      \label{mass}
\end{figure*}

Fig.~\ref{smc} shows the population synthesis calculations for reduced stellar metallicity $Z=0.002$ corresponding, for instance, to the SMC. As already discussed in section~\ref{SectionSMCLMC} at this low metallicity
the stellar evolution tracks spend most of their time in the group 1 BSG phase and group 2 becomes unimportant. In consequence, we see only a group 1 sequence predicted by the theory. For 
evolutionary models without rotation this sequence is very similar to the solar metallicity $Z=0.014$ group 1 sequence and, thus, also disagrees with the observed FGLR sequence. On the other hand, 
as for models with solar metallicity, the sequence produced by the models including the effects of rotation is in good agreement with the observations.

The main result of the above discussion is that models accounting for rotation do a better job for reproducing the observed FGLR relation provided most of the observed BSG belong to group 1.
Interestingly, we can check this conclusion independently by comparing predicted evolutionary masses with observed spectroscopic masses. First, we can test whether rotating
models indeed reproduce the spectroscopic masses better and, second, whether population synthesis without group 2 BSG leads to a better agreement at $Z = 0.014$. These comparisons are shown in  Fig.~\ref{mass}, where a typical error of $10\,M_\odot$ has been assumed.

In the range $1.9 \geq \log g/T_\text{eff,4} \geq 1.3$ the models including the effects of rotation indeed produce a probability ridge line, which agrees well with the observations, whereas models
without rotation predict too high masses. For $\log {g/T_{\rm eff,4}} < 1.3$, however, both sets of models seem to fail with too low predicted masses. As demonstrated by the lower left panel of the figure, this is due to the group 2 BSGs.

The lower right panel of Fig.~\ref{mass} displays the population synthesis with low metallicity models at $Z=0.002$, which include rotation. In this case, group 2 BSGs were not removed, since
for reasons already explained above the evolution models predict only a small number of group 2 BSG with masses not much lower than group 1. Most importantly, the high density probability ridge line
is again in good agreement with the observations. The comparison of predicted stellar evolution masses with observed spectroscopic masses, thus, corroborates the conclusions obtained from the comparison of observed and model FGLRs.
 
Finally, we note the presence of three high mass stars among the observed BSGs at $\log {g/T_{\rm eff,4}} \approx 1.25$ which  do not seem to agree with the trend predicted by the models and which also appear to be somewhat separated from the other BSG observed. In the FGLR-plane, these three points are also slightly above the bulk of the other observed points. It is tempting to explain this discrepancy by unresolved binarity or multiplicity. However, while this never can be ruled out completely, it would require two objects of very similar effective temperature and gravity and roughly equal luminosity to produce a spectrum of one single BSG. Given the short evolutionary lifetime of BSGs we regard this as unlikely. We also note that the objects are in a domain where the probability for their existence is not completely unlikely. Thus, their existence could just be a by chance effect of small numbers. 

We tested, using the models by \citet{Brott2011} and \citet{Georgy2013}, whether assuming a range in initial rotational velocities could affect the comparison between stellar and evolutionary masses.  We concluded that the observed scatter cannot be explained by 
considering a range in initial rotational velocities and we obtained a confirmation that rotation tends to improve the agreement between evolutionary and spectroscopic masses.

%{\color{red}{\bf In Fig.~\ref{massvit}, we test whether assuming a range in initial rotational velocities could affect the comparison between stellar and evolutionary masses. For this purpose, we use the models by \citet{Brott2011} and \citet{Georgy2013}. We see that the shift of the location of models is small if we restrict the range of initial velocities
%between 0 and 431 km s$^{-1}$. The shift is enlarged when initially very fast rotating stars are added, {\it i.e.} stars with initial rotations between 431 and 542 km s$^{-1}$ leading to an effect 
%mostly at lower gravities. However, as argued already above the likelihood to have successors of such stars among the observed BSG is very small. Thus, Fig.~\ref{massvit} indicates that the 
%conclusion from above still holds when a larger distribution of initial velocities is considered. It confirms also that rotation tends to improve the agreement between evolutionary and spectroscopic masses.
%}}

%\begin{figure}
%\centering
%{\bf 
%\includegraphics[width=.48\textwidth, angle=0]{brovitmc.pdf} 
%   \caption{Evolutionary masses compared with observed BSG spectroscopic masses. {\color{red}{\bf The lines show the same evolutionary models as in Fig.~7.evolutionary models. The upper (magenta) 
%curve corresponds to the non-rotating models, the (magenta) middle one to the rotating models with an initial rotation between 431 and 475 km $s^{-1}$, the lowest (blue) line to rotating stellar modelwith an initial velocities between 542 and 574 km $s^{-1}$.}}}
%      \label{massvit}
%}      
%\end{figure}

\section{Conclusions}

Summarizing the results from the previous sections we conclude that 
\begin{itemize}
\item the observed FGLR sequence of blue supergiant stars is reproduced best by stellar evolution models which include the 
effects of rotation and represent the evolutionary phase when the BSG evolve towards the red supergiant stage (group 1). 
\item The group 1 stellar model FGLR sequence is not affected by variations in metallicity. This provides strong support for the FGLR as a tool to determine extragalactic distances.
\item Models without rotation generate a group 1 FGLR sequence which is too bright when compared with the observations. 
\item At solar metallicity stellar models also predict the existence of group 2 BSG representing the evolutionary stage returning from the red supergiant stage with flux-weighted gravities distinct 
from group 1 and in much larger numbers. The difference in flux-weighted gravity of group 2 is caused by mass-loss in the red supergiant stage. The observations do not support the existence of 
group 2 or a difference with respect to group 1 in flux-weighted gravity.
\end{itemize}

While the overall good representation of the observed FGLR by the stellar models can be regarded as a success of stellar evolution theory, the disagreement with regard to group 2 at solar
metallicity poses a problem, which leads to three {possibilities for the post main-sequence evolution scenarios} of solar metallicity stars
with initial masses between $12$ and $40\,M_\odot$:
\begin{itemize}
\item Possibility 1: stars cross the Hertzsprung-Russel diagram and end their evolution as RSG. In that case, no group 2 blue supergiants are predicted.
\item Possibility 2: stars cross the Hertzsprung-Russel diagram, go through a RSG phase, start to evolve back to the blue but explode as core collapse supernova either prior to the group 2 phase or 
soon after entering it. In that case, group 2 blue supergiants have such short lifetimes that the probability to observe them is
very small. 
\item Possibility 3: stars evolve back to the blue after a red supergiant phase but without losing too large amounts of mass during the RSG phase.
The physics responsible for the blueward evolution after the RSG phase is the development of a relatively massive helium core which drives the star back to the helium burning sequence which for 
pure helium stars is at 
effective temperatures hotter than the hydrogen burning sequence. According to \citet{Gia1967} the minimum mass fraction of the helium core to evolve back should be 60-70\% of the total mass.
Such a high fraction can be obtained through strong mass-loss but also through mixing processes. In this scenario, strong mixing processes would drive the star back with $M/L$ ratios not very
much different from group 1.
 \end{itemize} 

The three possibilities are not exclusive in the sense that they may all occur either in different initial mass regimes or for the same initial mass but presenting an additional difference caused, 
for instance, by rotation and/or the presence of a companion star. What does appear to be excluded or to be an infrequent scenario is the one in which the stars evolve into the RSG stage, 
lose large amounts of mass and then spend a significant fraction of the core He-burning lifetime as group 2 BSG. 

The three possibilities above predict different outcomes which could be tested by observations.
If possibility one is the most frequent, then the number of stars ending their lifetimes as a blue or yellow supergiant should be very rare. On the other hand,
if possibility two is the most frequent, then many supernovae should explode inside material very recently released by the last strong mass loss episode. This may lead to narrow 
absorption lines superimposed to the supernova spectra and caused by the low velocity surrounding stellar wind material.
Possibility 3 should produce a significant amount of group 2 blue supergiants. While the stellar gravities of these objects would be comparable to group 1, the chemical composition will be very
likely different because of the strong effects of mixing during the RSG stage.

While tests of possibilities 1 and 2 require either extended surveys for the progenitors of core collapse supernovae or high resolution, high signal-to-noise spectroscopy of such supernovae, an 
investigation of possibility 3 by discriminating between group 1 and group 2 blue supergiants appears to be more straightforward.
The surface composition of group 2 BSG should present stronger signs of CNO processing than group 1. A comprehensive project studying the surface composition of solar metallicity BSG
by means of high resolution, high signal-to-noise spectroscopy could provide a way to distinguish between the pre- and post RSG blue supergiants. However, the mixing induced by rotation during the 
main sequence phase can blur the picture somewhat, since depending on the initial rotation rate, significant changes of the surface abundances can already be obtained at the end of the 
main sequence phase. Thus, surface abundances can result from either strong mixing during the main sequence or from a much less efficient mixing during the main sequence phase and a dredge up effect 
during a RSG stage. Moreover, close binary evolution may also deeply modify the surface composition of BSGs. The questions is made complicated by the fact that mergers may make the star to appear as single. It is however the hope that
further studies will find relations between surface rotation, surface abundances, and pulsation properties that will be specific signatures of the single, respectively close binary scenarios.
We also note that the abundances obtained at the surface of post red supergiant are sensitive to the way
convection is computed. Different results are obtained when the Schwarzschild or the Ledoux criterion for convection is used \citep{Georgy2014}. Still, accurate spectroscopy will very likely reveal
whether scenario three is likely or not.

Another approach to observationally investigate the three possible scenarios is a detailed and comprehensive study of BSG photometry. \citet{Saio2013} predict that group 2 BSGs have very different pulsational properties compared to group 1. It would, therefore, be possible to determine the fraction of group 1 and group 2 BSGs observationally by a careful analysis of their variability.

Independent of the problem of how to explain the apparent absence of group 2 objects in the observed FGLR sequence of BSGs our investigation provides an improved foundation of the use of the FGLR-method
for the determination of extragalactic distances. A major remaining uncertainty of the application of this method so far has been the potential metallicity dependence of this relationship, although 
the spectroscopic work by \citet{Urba2008} and \citet{Hosek2014} indicated that metallicity effects should be small. The population synthesis results obtained in this study based on most recent 
stellar evolution models support this conclusion establishing the FGLR-method as a robust and accurate way to determine distances to galaxies. A needed follow-up of the present work will be the study of the impact of close binary evolution on the small scatter of the FWGL relation.

\begin{acknowledgements}
This work was supported by the Swiss National Science Foundation (project number 200020-146401) to GM, the National Science Foundation under grants AST-1008798 and AST-1108906 to RPK,
and by the European Research Council under the European Union Seventh Framework Programme (FP/2007-2013) / ERC Grant Agreement n. 306901. 
GM gratefully acknowledges the hospitality of the  Institute for Astronomy, University of Hawaii, where part of this work was carried out. 
\end{acknowledgements}

\bibliographystyle{aa}
\bibliography{ms_rpk_corr.bib}
%\bibliography{ms}

\end{document}